\documentclass[journal, onecolumn]{IEEEtran}
\IEEEoverridecommandlockouts
\usepackage{cite}

\usepackage{dcolumn}% Align table columns on decimal point

\usepackage{bm}
\usepackage{booktabs}
\usepackage{multirow}
\usepackage{algorithm}
\usepackage{algpseudocode}
\usepackage{subfigure}
\usepackage{amssymb}
\usepackage{cite}
\usepackage{multicol}
\usepackage{lipsum}% dummy text
\usepackage{pstricks}
\usepackage{array}
\usepackage[utf8]{inputenc}
\usepackage[T1]{fontenc}
\usepackage{url}
\usepackage{ifthen}
\usepackage{cite}
\usepackage{color, xcolor}
\usepackage[cmex10]{amsmath}
\usepackage{arydshln}
\usepackage{graphicx}
\usepackage{framed}
\usepackage{epstopdf}
\usepackage{amsmath}
\usepackage{amsthm}
\usepackage{amsfonts}
\usepackage[justification=centering]{caption}
\usepackage{soul} % µŒÈë soul °ü
\usepackage{color, xcolor}
\theoremstyle{}
\newtheorem{theorem}{Theorem}
\newtheorem{lemma}{Lemma}

\newtheorem{proposition}{Proposition}

\newtheorem{remark}{Remark}

\begin{document}

\title{A Novel Coded Computing Approach for Distributed Multi-Task Learning}

\author{Minquan Cheng, Yongkang Wang, Lingyu Zhang, and Youlong Wu

%\thanks{Manuscript received April 19, 2021; revised August 16, 2021.}}
\thanks{ M. Cheng, Y. Wang, and L. Zhang are with Guangxi Key Lab of Multi-source Information Mining $\&$ Security, Guangxi Normal University,
Guilin 541004, China  (e-mail: chengqinshi@hotmail.com, zcyongkang@stu.gxnu.edu.cn, lingyu.zhang128@stu.gxnu.edu.cn.).
}
\thanks{Y. Wu is with the School of Information Science and Technology, ShanghaiTech
University, Shanghai 201210, China. (e-mail: wuyl1@shanghaitech.edu.cn). }
}

\maketitle

\begin{abstract}
Distributed multi-task learning (DMTL) effectively improves model generalization performance through the collaborative training of multiple related models. However, in large-scale learning scenarios, communication bottlenecks severely limit practical system performance. In this paper, we investigate the communication bottleneck within a typical DMTL system that employs non-linear global updates. This system involves distributed workers, assisted by a central server, who collaboratively learn distinct models derived from a non-linear aggregation of their local model parameters. We first characterize the communication process as a matrix decomposition problem. It transforms workers' data storage constraints into structural characteristics of the uplink encoding matrix, and worker data retrieval demands into Maximum Distance Separable (MDS)  properties of the downlink encoding matrix. Building on this, we propose a novel coded DTML scheme that can greatly reduce the communication cost of the DTML with heterogeneous data placement. Theoretical analysis demonstrates that the proposed scheme achieves the theoretical lower bound for communication overhead under mild conditions. Remarkably, this optimality holds for both traditional homogeneous computing environments and various heterogeneous scenarios. Furthermore, our scheme is extensible to a distributed linearly separable computation problem where the target function involves multiple linear combinations of local update values. This indicates that our scheme offers a new way of tackling heterogeneous data placement challenges in various distributed applications.

\end{abstract}


\section{Introduction}
  Distributed computing has emerged as a pervasive paradigm for addressing the challenges of large-scale data processing and the resolution of complex computational problems.  By exploiting the distributed computing and storage resources across multiple interconnected nodes, distributed computing systems offer enhanced scalability, fault tolerance, and processing power, and thus have been widely used in big data analytics, machine learning,  cloud services, etc.    However, in large-scale distributed environments, information exchange among the distributed nodes (e.g., gradient transmission, global model transmission) incurs high communication delays due to heavy communication loads and limited resources (e.g., communication bandwidth, transmission power). This leads to serious communication bottlenecks, which in turn limit the efficiency and scalability of distributed computing systems \cite{2020Communication,2022Edge}.

To mitigate the aforementioned communication bottlenecks, coding techniques have been introduced into distributed computing to reduce the impact of limited communication resources. In particular, the authors in \cite{LMMYA} proposed a coded distributed computing (CDC) scheme based on coded caching technique \cite{MN} for the MapReduce framework  \cite{DG}, in which redundant computation and coded multicasting are exploited to achieve the optimal computation-communication trade-off and reduce the communication overhead.  The CDC approach has garnered wide research interest recently, as evidenced by a multitude of studies investigating its various facets \cite{li2017scalable, li2019wireless, bi2022dof, wu2023coded, chen2021coded, chen2023optimality, HWSLJZ, hu2023coded, woolsey2021new, xu2021new, reisizadeh2019coded, wang2022coded, song2022joint, JQ, cheng2023asymptotically, WCJ}.  For instance, early work by \cite{li2017scalable} introduced a centralized CDC scheme (CWDC), where workers collaboratively execute MapReduce tasks and exchange intermediate values through a central server. Expanding on this, \cite{li2019wireless, bi2022dof, wu2023coded} further explored wireless interference in distributed computing, leveraging techniques like zero forcing \cite{li2019wireless} and interference alignment \cite{bi2022dof, wu2023coded} to achieve multiplex gains. Additionally, \cite{chen2021coded, chen2023optimality} proposed a master-aided CDC scheme, where a central server assists workers in MapReduce tasks by performing Map operations. To address the exponential increase in computation complexity of CDC schemes growing exponentially with the number of computing nodes, \cite{JQ, cheng2023asymptotically, WCJ} introduced low-complexity schemes utilizing combinatorial designs. Woolsey et al. \cite{woolsey2021new} studied the schemes for both homogeneous and heterogeneous distributed computing networks; Xu et al. \cite{xu2021new} characterized the upper and lower bounds of the optimal communication load as two linear programming problems for a general heterogeneous CDC system.

The MapReduce framework, although revolutionary for batch processing of large datasets, is generally not well-suited for distributed learning due to inherent limitations on processing iterative algorithms \cite{NIPS2014_935ad074}.  In a typical distributed learning,  distributed workers compute local gradients using their local data batches (namely local update), and then send them to a parameter server. After gathering all local updates, the parameter server aggregates local updates (namely global update) and broadcasts the global update model to all workers for the next learning iteration.  To improve the communiation efficiency and robustness against stragglers,  a {gradient coding} scheme was proposed by Tandon, et al. in \cite{TLDK} for the single-task distributed learning frameworks, where the global update is just a linear aggregation of the local gradients; Subsequently, Ye et al. in \cite{2018Communication} introduced a communication-efficienct gradients coding scheme to achieve an optimal tradeoff among the computation load, straggler tolerance, and communication cost; Building upon this, the authors \cite{2022Adaptive} proposed an adaptive gradient coding scheme that achieves the optimal communication cost for changing number of stragglers.  Wan et al. \cite{2022Distributed} investigated a more general problem--distributed linearly separable computation (LSC), where the global update can be expressed as multiple linear combinations, and proposed a zero forcing-based scheme with the minimum communication costs. The authors later in \cite{2023Fundamental} proposed a scheme based on interference alignment to further reduce the communication cost.  Note that the aforementioned works \cite{TLDK,2018Communication,2022Adaptive,2022Distributed,2023Fundamental} mainly focus on a homogenous setting (equal amount of data batches assigned to all workers) and linear aggregation (i.e., the global update performs linear combination of local updates). Recent works in \cite{Tang2021, HWSLJZ} considered a distributed multi-task learning (MTL)  setting \cite{zhang2021survey,liu2017distributed,smith2017federated,Weinberger'09}, where workers are assigned \emph{unequal} numbers of data batches, and the global update performs \emph{nonlinear} aggregation. In particular,   \cite{HWSLJZ} first introduced coded MTL schemes that can greatly reduce downlink and uplink communication costs, and later \cite{HLCMSW} extended coding techniques to the hierarchical MTL system.  Unfortunately, these works can only achieve the optimal communication-computation trade-off under a specific symmetric setting with equal storage and computation cost across all distributed workers. The achievability of theoretical lower bounds under the heterogeneous setting and more relaxed homogeneous conditions remains an open and challenging problem.

In this paper, we revisit the heterogeneous distributed MTL problem where a central server is connected to $K$ workers with random data assignments, as shown in Fig. \ref{fig.1}. We aim to exploit the structure of DMTL to achieve both theoretically optimal uplink and downlink communication loads. The main contributions are summarized as follows.
\begin{itemize}

\item We characterize the encoding and decoding processes in heterogeneous coded MTL as a matrix decomposition problem. This involves equivalently transforming data storage constraints into structural characteristics of the uplink encoding matrix, and worker data retrieval demands into Maximum Distance Separable (MDS)  properties of the downlink encoding matrix. This specific transformation has not been utilized in prior works \cite{HWSLJZ,HLCMSW}, and paves the way for designing optimal coded computation schemes.
\item Based on our matrix decomposition model, we developed a novel coded  MTL  scheme.  Compared to state-of-the-art schemes in \cite{HWSLJZ,HLCMSW}, our method significantly reduces both uplink and downlink communication loads.
\item We establish the sufficient condition for our scheme's optimality, by leveraging the MDS property and Hall’s Marriage Theorem \cite{P.C}. This condition, which depends solely on workers' data placement, is remarkably mild. It encompasses both the homogeneous scenarios in \cite{HWSLJZ,HLCMSW} and a wide range of heterogeneous scenarios.

\item Interestingly, we found that the proposed scheme is also applicable to the LSC problem \cite{2022Distributed, 2023Fundamental}, and can handle more complex heterogeneous data placements.  This extension is notable because existing schemes in \cite{2022Distributed, 2023Fundamental} are typically limited to the homogeneous setting with cyclic data placement.  This limitation primarily stems from their reliance on the Schwartz-Zippel lemma for feasibility proofs, which becomes exceptionally challenging in heterogeneous settings.
\end{itemize}

 The rest of this paper is organized as follows: Section \ref{sec:system model-description} introduces the DMTL system model and reviews existing results. Section \ref{Main Results} presents the optimality results, provides their numerical result analysis, and an example of the proposed scheme. The proof of the key theorem is detailed in Section \ref{Scheme}. Finally, Section \ref{CONCLUSION} concludes the paper.

In this paper,  we use the following notations unless otherwise stated.
We use $\mathbb{N}^+$ and $\mathbb{R}$ to denote the positive integers and real numbers, respectively.
$\mathbb{F}_{q}$ denotes a finite field of order $q$, where $q$ is a power of a prime number.
$|\cdot|$ is used to represent the cardinality of a set or the length of a vector.
For any positive integers $a$, $b$ with $a<b$, let $[a : b]=\{a,a+1,\ldots,b\}$, especially $[1 : b]$ is shortened to $[b]$.
For a matrix $\mathbf{M}\in \mathbb{F}_{q}^{k\times n}$, then $\mathbf{M}(i,j), i \in [k], j\in [n]$, represents the entry in the $i$th row and $j$th column. $\mathbf{M} (i, [n]), i\in [k]$ represents the $i$th row, and $\mathbf{M}([k], j), j\in [n]$ represents the $j$th column of matrix $\mathbf{M}$.

\section{Problem Formulations}\label{sec:system model-description}
In this section, we first introduce the coded DMTL problem model and then present existing results in the literature.
\subsection{The Distributed Multi-Task Learning System}
\begin{figure}[t]
\centering
\includegraphics[scale=0.45]{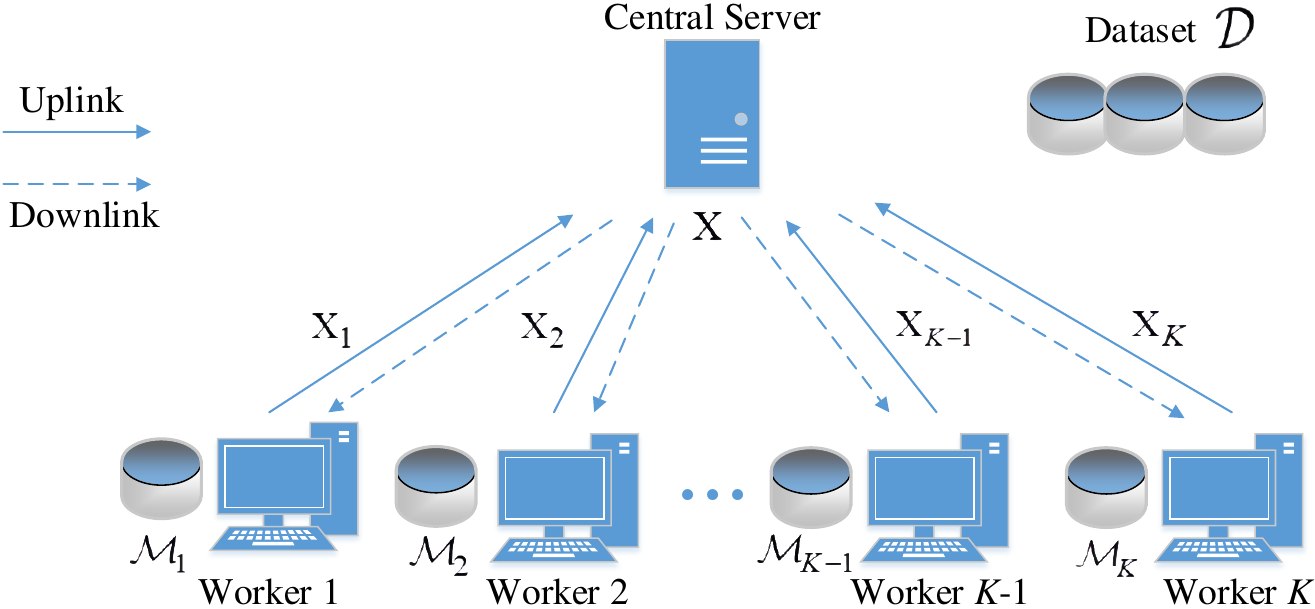}
\caption{ A DMTL system consists of a central server connected to $K$ distributed workers via an error-free shared link. During each iteration, each worker $k\in [K]$ obtains a local update $\mathcal{V}_k$ based on local data and model, and then sends a message $\mathbf{X}_{k}$ to the server via the uplink, and the server sends a message $\mathbf{X}$ to the workers via the downlink, such that each worker $k$ obtains the updated $\omega_k$.}
\label{fig.1}
\end{figure}

Consider a $(K,N;\{\mathcal{Z}_k\ | \ k\in[K]\})$ DMTL system, as illustrated in Fig. \ref{fig.1}.  This system comprises a central server, a dataset $\mathcal{D}$ partitioned into $N$ equal-sized batches (i.e., $\mathcal{D}=\{\mathcal{D}_1,\mathcal{D}_2,\ldots,\mathcal{D}_{N}\}$), and $K$ distributed workers. Each worker $k\in[K]$ stores a subset of dataset $\mathcal{M}_k=\{\mathcal{D}_n\ |\ n\in\mathcal{Z}_k\}$, where  $\mathcal{Z}_k$ is the index set of data batches managed by worker $k\in[K]$.

The $K$ workers connect to the central server via an error-free shared link, collaboratively aiming to learn their personalized models $\{\omega^{(t)}_k\in \mathbb{R}^m\}$. Here  $t$ and $m$ are positive integers denoting the iteration index and the model length, respectively.  Following similar setups in  \cite{HWSLJZ,HLCMSW}, we assume each worker $k$ wishes to learn a unique mode (i.e., $\omega^{(t)}_k\neq \omega^{(t)}_j$ for any integers $k\neq j$ and $t$).

A DMTL algorithm involves multiple iterations.  At iteration $t\geq 1$, workers perform a local update based on the local information ($\mathcal{M}_k$ and stored models)  and then send local update values (e.g., partial gradients) to the server via the uplink. Subsequently, the server broadcasts signals to all workers, enabling each worker $k$ to obtain its desired update models $\omega^{(t+1)}_k$ generated through a global update (i.e., aggregates local update values). We now divide the DMTL setting into its uplink and downlink processes for detailed discussion. Since all iterations operate in the same way, we will omit the superscript $(t)$ to simplify notations.
\begin{itemize}
\item \textbf{Uplink Process}: Given the local data batches $\mathcal{M}_k=\{\mathcal{D}_n\ |\ n\in\mathcal{Z}_k\}$, each worker $k\in[K]$ performs local update to computes the local update value  $\mathbf{v}_i\in\mathbb{F}_q^{E}$, say partial gradient,   from the dataset $\mathcal{D}_i$,  for every $i\in\mathcal{Z}_k $. Here  $E\in\mathbb{N}^+$ is the length of a partial gradient.  %has a set of data batches indexed by  $\mathcal{Z}_k$. It  first computes the local update value, say partial gradient,  based on the  local data and model $\omega_k$ obtained from the previous iteration, i.e., 
%\begin{align}
%\label{local update value}\mathbf{v}_i=f_{i}(\mathcal{D}_i,\omega_k)\in  \mathbb{F}_q^{E},~\text{for~all}~  i\in\mathcal{Z}_k,
%\end{align}
%where $E\in\mathbb{N}^+$ is the length of each partial gradient and  $f_i: \mathbb{R}^{\kappa}\times\mathbb{F}_q^{m}\rightarrow \mathbb{F}_q^{E}$ is a local update function of the $i$-th task. 
The local update function, for example, can be set as the square loss or hinge loss for Support Vector Machine (SVM) models \cite{smith2017federated}.  

As each worker computes $|\mathcal{Z}_k|$ data  batches, we define $r_k:=|\mathcal{Z}_k|$ as the {computation load} of worker $k\in[K]$, and the (average) computation load as
\begin{IEEEeqnarray}{rCl}
r:=\frac{\sum_{k\in [K]}r_k}{K}.
\end{IEEEeqnarray}
In this work, we consider a general scenario where workers randomly store data batches without central server coordination, implying a heterogeneous case where computation loads may vary among workers.

Let $\mathcal{V}_k$ be the set of local update available at worker $k$:
\begin{align}
\label{local update}
\mathcal{V}_k=\{\mathbf{v}_i\ |\  i\in\mathcal{Z}_k\}.
\end{align}

During the uplink communication, each worker $k\in[K]$  generates and transmits a coded message
\begin{align}
\label{uplink user}
\mathbf{X}_k=\psi_{k}(\mathcal{V}_k)\in\mathbb{F}_q^{\mathrm{T}_k}
\end{align}
to the server through an error-free shared link, where  $\psi_k:(\mathbb{F}_{q}^E)^{r_k}\rightarrow\mathbb{F}_q^{T_{k}}$ is the encoding function of worker $k$ for some positive integer $T_{k}$.
\item \textbf{Downlink Process}:
Define the global update as:
\begin{align}
\label{global update}
(\omega_1,\omega_2,\cdots,\omega_K)=\phi(\mathbf{v}_1,\mathbf{v}_2,\cdots,\mathbf{v}_N),
\end{align}
where $\phi: (\mathbb{F}_q^{E})^{N}\rightarrow(\mathbb{F}_q^{E})^{N}$ is the global update function. To enhance optimization effectiveness, non-linear global aggregations have been widely applied in DMTL. For instance, \cite{smith2017federated} proposed a non-linear global aggregation by learning a task relationship matrix that dictates non-linear dependencies between task models, rather than just linear combinations. Similarly,
\cite{MTLAttention} employed attention mechanisms to dynamically weigh task contribution by considering the rate of change of loss for each task. Consistent with approaches like \cite{HWSLJZ}, we assume that distributed workers perform the global update, allowing the local information at workers to be exploited as side information to reduce communication costs.

Based on messages $\mathbf{X}_1$,  $\mathbf{X}_2$, $\ldots$, $\mathbf{X}_K$ sent from the uplink, the server computes and broadcasts a coded message
\begin{align}
\label{downlink server}
\mathbf{X}=\psi(\mathbf{X}_1,\mathbf{X}_2,\cdots,\mathbf{X}_K)\in \mathbb{F}_q^{T}
\end{align}
for some positive integer $T$ and  $\psi:\mathbb{F}_q^{T_1}\times\mathbb{F}_q^{T_2}\times\cdots\times\mathbb{F}_q^{T_K}\rightarrow\mathbb{F}_q^{T}$ is the encoding function at the server. After receiving the message $\mathbf{X}$, each worker $k$ utilizes its decoding function $\varphi_k: \mathbb{F}_q^{T}\times(\mathbb{F}_q^{E})^{r_k} \rightarrow(\mathbb{F}_q^{E})^{K}$ to recover $(\mathbf{v}_1,\mathbf{v}_2,\cdots,\mathbf{v}_N)$, specifically
\begin{align}
\label{recover IVs}
(\mathbf{v}_1,\mathbf{v}_2,\cdots,\mathbf{v}_N)=\varphi_k(\mathbf{X}, \mathcal{V}_k).
\end{align}Finally, the workers compute global update \eqref{global update} from the set $\{\mathbf{v}_n\}_{n\in [N]}$  to obtain the updated model parameters.

\end{itemize}
Given the aforementioned processes, the normalized uplink and downlink communication loads are defined as
\begin{align}
\label{eq-load-def}
L_{\text{up}}\triangleq\frac{\sum_{k\in[K]}T_k}{E}~ \text{and}~L_{\text{down}}\triangleq \frac{T}{E},
\end{align} respectively. The goal is to design a $(K,N;\{\mathcal{Z}_k\ | \ k\in[K]\})$ DMTL scheme that achieves both the minimum uplink communication load  $L_{\text{up}}^\ast$ and the minimum downlink communication load $L_{\text{down}}^\ast$.
\subsection{Existing Results}
For each data subset $\mathcal{D}_n$ where $n\in[N]$, let $\mathcal{K}_n$ represent the set of workers storing  $\mathcal{D}_n$, i.e., $\mathcal{K}_n=\{k\in [K]\ |\  n\in \mathcal{Z}_k\}$. For any integer $z\in[K]$, we define
$\mathcal{I}_z$ as the set of data batches, each of which is stored exactly $z$ times, i.e., $\mathcal{I}_z=\{n\in[N]\ |\ |\mathcal{K}_n|=z\}$.

 For any $(K,N;\{\mathcal{Z}_k\ | \ k\in[K]\})$ DMTL problem, the authors in \cite{HWSLJZ} established the following converse and achievable bounds on the communication loads.
\begin{lemma}[Lower bounds on  Communication Loads]\rm\cite{HWSLJZ}\label{lem-HWSLJZ-converse}
Given a $(K,N=K;\{\mathcal{Z}_k\ | \ k\in[K]\})$ DMTL scheme, the converses of the uplink and downlink communication loads are
\begin{align}
\label{eq-HWSLJZ-converse}
L_{\text{up}}^\ast  \geq \frac{KN}{K-1}-\sum_{z\in[K]}\frac{z|\mathcal{I}_z|}{K-1}
\ \ \ \ \text{and} \ \ \ \ L_{\text{down}}^\ast  \geq N-\min_{k\in [K]}|\mathcal{Z}_k|,
\end{align}respectively.
\hfill $\square$
\end{lemma}
\begin{lemma}[ Achievable Scheme in \cite{HWSLJZ}]\rm
\label{lem-HWSLJZ-scheme}
For any positive integers $K$, and $K$ subsets $\{\mathcal{Z}_k\ | \ k\in[K]\}$ of $[K]$, there is a  $(K,N=K;\{\mathcal{Z}_k\ | \ k\in[K]\})$ DMTL scheme with the uplink and downlink communication loads
\begin{align}
\label{eq-HWSLJZ-scheme}
L_{\text{up}}=N-\sum_{z\in [K]}\sum_{k\in[K]}\frac{1}{z}\min_{j\in [K]}\mid\mathcal{Z}_k\cap \mathcal{Z}_j\cap\mathcal{I}_z\mid \ \ \ \text{and} \ \ \  L_{\text{down}}=N-\sum_{z\in[K]}\min_{k\in [K]}|\mathcal{Z}_k\cap \mathcal{I}_z|,
\end{align} respectively.\hfill $\square$
\end{lemma}

In \cite{HLCMSW}, the authors proposed a novel coded scheme for the hierarchical MTL system where a central server connects with workers via intermediate relay nodes. Their scheme exploits the underlying network topology to avoid unintended data sharing and information leakage among the workers.   With minor modifications, specifically by interpreting the relay nodes as workers and the relay-user network topology as the data placement strategy, this hierarchical coded MTL scheme is extensible to the conventional MTL setting. The resulting communication loads are provided as follows.
\begin{lemma}[Achivable Scheme in \cite{HLCMSW}]\rm
\label{lem-HLCMSW-scheme}
For any positive integers $K$, $N$, and $K$ subsets $\{\mathcal{Z}_k\ | \ k\in[K]\}$ of $[K]$, there is a  $(K, N; \{\mathcal{Z}_k\ | \ k\in[K]\})$ DMTL scheme with the uplink and downlink communication loads
\begin{align}
\label{eq-HLCMSW-scheme}
L_{\text{up}}=\sum_{z\in [K-1]}\sum_{k\in [K]}\alpha_k^z(|\mathcal{Z}_k\cap \mathcal{I}_z|-\min_{j\in [K]}|\mathcal{Z}_k\cap\mathcal{Z}_j \cap\mathcal{I}_z|) \ \ \ \text{and} \ \ \
L_{\text{down}}=N-\sum_{z\in [K-1]}\min_{k\in [K]}|\mathcal{Z}_k\cap \mathcal{I}_z|,
\end{align} where the coefficients $\{\alpha_{k}^{z}\}$ are determined by solving an optimization problem.
  \hfill $\square$
\end{lemma}

\begin{remark}\rm
\label{DMTL-HMTL} The achievable scheme in \cite{HLCMSW} always outperforms that in \cite{HWSLJZ}. This is because,
when $\alpha_{k}^{z} = \begin{cases}
\frac{1}{z} & \text{if } |\mathcal{Z}_{k} \cap \mathcal{I}_{z}| \neq 0 \\
0 & \text{if } |\mathcal{Z}_{k} \cap \mathcal{I}_{z}| = 0
\end{cases}$,
the load $L_{\text{up}}$ in \eqref{eq-HLCMSW-scheme} turns to
$$L_{\text{up}} \leq N-\sum_{z\in [K-1]}\sum_{k\in [K]}\frac{1}{z}\min_{j\in [K]}\mid\mathcal{Z}_k\cap \mathcal{Z}_j\cap\mathcal{I}_z\mid .$$ In this case, the result of Lemma \ref{lem-HLCMSW-scheme} is the same as that in Lemma \ref{lem-HWSLJZ-scheme}.
\end{remark}

It has been shown that the schemes in \cite{HLCMSW,HWSLJZ} achieve the minimum communication loads when  $r_1=r_2=\cdots=r_K=r$ and under a symmetric data placement.
However,  for scenarios involving unequal computation loads, or even with equal computation loads but asymmetric data placement, the achievability of theoretical lower bounds remains an open question. In this paper, we aim to design schemes that achieve the optimum for both uplink and downlink communication loads of the DMTL under \emph{random} data placement.

\section{Main Results}\label{Main Results}
In this section, we first present our optimality result, followed by its numerical result analysis, and finally, an example of the proposed scheme.
\subsection{Optimality Results}
Given a $(K,N;\{\mathcal{Z}_k\ | \ k\in[K]\})$ DMTL problem, recall that we assume $r_k=|\mathcal{Z}_k|$ for each integer $k\in[K]$. We divide each local update value $\mathbf{v}_n$ into $K-1$ packets. To formally introduce our main result, the following notations are essential. For each integer $k\in[K]$, we define
\begin{align}
\label{eq-key-parameters}
d_k=N+(K-1)r_k-\sum_{k'\in[K]}r_{k'},\ \ \ S=\sum_{k\in[K]}d_k.
\end{align}
As detailed in Proposition \ref{pro-1},   $d_k$ represents the minimum number of coded packets transmitted by worker $k$, while $S$ denotes the minimum total number of coded packets transmitted by all workers. For each integer $k\in[K]$, let $\mathcal{P}_k$ represent the index subset of the packets each of which is not stored by worker $k$, i.e.,
\begin{align}
	\label{eq-combinatoric}
	\mathcal{P}_k = \bigcup_{i \in [0 : K-2]} \left\{ ([N] \setminus \mathcal{Z}_k) + iN \right\}.
\end{align}
In this paper, we obtain our main result under the following condition
\begin{center}
{\bf Condition 1:} $| \bigcap_{i\in[k]}\mathcal{P}_i |\leq S-\sum_{i\in[k]}d_i$ holds for any positive integer $k\in[K]$.	
\end{center}
This is achieved by first introducing a novel matrix decomposition that unifies the characterization of the DMTL problem, and then by exploiting the MDS property and applying Hall's Marriage Theorem \cite{P.C} to generate the optimal scheme. The resulting optimality is formalized in the following theorem, whose proof is provided in Section \ref{Scheme}.
\begin{theorem}[Optimal scheme]\rm
	\label{th-1}
	In a $(K,N;\{\mathcal{Z}_k |  k\in[K]\})$ DMTL system satisfying {\bf Condition 1}, there exists a $(K,N;\{\mathcal{Z}_k\ | \ k\in[K]\})$ DMTL scheme with minimum uplink and downlink communication loads, i.e., $L_{\text{up}}^\ast = \frac{KN}{K-1}-\sum_{z\in[K]}\frac{z|\mathcal{I}_z|}{K-1}$ and $L_{\text{down}}^\ast = N-\min_{k\in [K]}|\mathcal{Z}_k|$.
	\hfill $\square$
\end{theorem}
\begin{remark}[Key idea of our optimal scheme]\rm
\label{remark-key-idea}
In this paper, we use the idea of matrix decomposition $\mathbf{P}=\mathbf{A}\mathbf{B}$ to characterize the encoding and decoding processes of the DMTL scheme. Here  $\mathbf{P}$ is called the uplink matrix in Subsection \ref{subsect-uplink}, representing the placement state and uplink communication strategy simultaneously, and $\mathbf{B}$ is called the downlink matrix in Subsection \ref{subsect-down}, representing the downlink communication strategy. From this perspective, the key idea of our optimal scheme is as follows:
\begin{itemize}
\item The minimum downlink load can be achieved if the downlink matrix has the MDS property. By coding theory, we can always construct $\mathbf{B}$ satisfying the MDS property when the operation field is large. In addition, the MDS property enables us to unify the uplink and downlink processes of the scheme, and can achieve the minimum uplink and downlink loads using a matrix equality.
\item The minimum uplink load can be achieved if there exists an invertible matrix $\mathbf{A}$. It is worth noting that proving the invertibility of matrix $\mathbf{A}$ is non-trivial due to the inherent randomness in data subset placement. This randomness introduces two primary challenges: (i) identifying the precise combinatorial condition {\bf Condition 1} for the random placement strategy that guarantees minimum uplink load, and (ii) leveraging the MDS property of matrix $\mathbf{B}$ to rigorously prove the linear independence of the rows of $\mathbf{A}_1$, $\mathbf{A}_2$, $\ldots$, $\mathbf{A}_K$ which make up $\mathbf{A}$ under {\bf Condition 1}. Our approach addresses these by establishing a relationship between the data placement and the intersection of subspaces spanned by the columns of $\mathbf{B}_k$ matrices, subsequently employing the MDS property of $\mathbf{B}$ and Hall’s Marriage Theorem \cite{P.C} to prove $\mathbf{A}$'s invertibility, as detailed in Subsection \ref{subsection-invertible-A}.
\end{itemize}
\end{remark}
\begin{remark}[Extension To Linearly Separable Computation Problem \cite{2022Distributed, 2023Fundamental}]\rm
\label{remark-extension}
We find that our scheme in Subsection \ref{Scheme} is also applicable to the distributed LSC problem  \cite{2022Distributed}. 	Recall that in the problem of distributed LSC, the server aims to compute a linear function $g$ that represents  $N_c$ linear combinations of $N\geq N_c$ local update values $\mathbf{v}_{1}$, $\ldots$, $\mathbf{v}_{N}$, i.e., $g(\mathbf{v}_{1},\mathbf{v}_{2},\ldots,\mathbf{v}_{N}) = \mathbf{F} (\mathbf{v}_{1} ;\cdots ;\mathbf{v}_{N})$,
where $\mathbf{F}$ is a $N_c \times N$ matrix known to both the server and the workers, whose elements are i.i.d. uniformly distributed over $\mathbb{F}_q$. By using our scheme with only the uplink process, and if we regard the uplink matrix $\mathbf{P}$ as the coefficient matrix for workers' uploaded information and set the submatrix $\mathbf{B}'$ of the downlink matrix $\mathbf{B}$ to $\mathbf{F}$, our scheme is feasible in the distributed LSC problem. This is feasible because $\mathbf{B}'$ has the MDS property, which ensures that the entries of $\mathbf{B}'$ are uniformly i.i.d. over the finite field $\mathbb{F}_q$. It is worth mentioning that the scheme in \cite{2022Distributed} is limited to cyclic data placement, whereas ours offers generalizability to heterogeneous data placement scenarios.
\end{remark}

\subsection{Numerical Results}

In this subsection, we compare the existing DMTL schemes with our scheme. Since the scheme in \cite{HLCMSW} does not have an explicit form on the communication loads (as seen in Lemma \ref{lem-HLCMSW-scheme}), we'll use numerical methods for a comparative analysis.

Without loss of generality, we use the DMTL scheme with parameters $(K=4, N=6, \{\mathcal{Z}_k \mid k \in [4]\})$ as an example for numerical comparison with the results in \cite{HLCMSW}. Regarding data placement, we start with a scenario where   $|\mathcal{Z}_1|=|\mathcal{Z}_2|=|\mathcal{Z}_3|=|\mathcal{Z}_4|=2$ (i.e., $r=2$), and randomly select a data placement that satisfies {\bf Condition 1}. For other data placement cases, we add data subsets to the above $r=2$ scenario and ensure that it still satisfies {\bf Condition 1}. The specific data placement scheme is as follows.
\begin{center}\label{table 1}
	{\textbf{Table  \textrm{I}:} Data Placement}\\
		\vspace{1em}
	\begin{tabular}{c|c|cccc}
  \hline
  $\sum_{k\in[4]}{r_k}$  & $r$ & $\mathcal{Z}_1$  & $\mathcal{Z}_2$ & $\mathcal{Z}_3$ & $\mathcal{Z}_4$\\ \hline
            $8$          &  2  &    $\{1,2\}$     &      $\{1,4\}$  &    $\{2,6\}$    & $\{3,5\}$ \\
  $9$ &$9/4$& $\{1,2,3\}$ & $\{1,4\}$ & $\{2,6\}$ & $\{3,5\}$\\
 $10$ &$5/2$& $\{1,2,3\}$ & $\{1,4,5\}$ & $\{2,6\}$ & $\{3,5\}$ \\
 $11$ &$11/4$& $\{1,2,3\}$ & $\{1,4,5\}$ & $\{2,4,6\}$ & $\{3,5\}$\\
$12$ &$3$& $\{1,2,3\}$ & $\{1,4,5\}$ & $\{2,4,6\}$ & $\{3,5,6\}$\\
 $13$& $13/4$& $\{1,2,3\}$ & $\{1,4,5\}$ & $\{1,2,4,6\}$ & $\{3,5,6\}$\\
 $14$& $7/2$& $\{1,2,3\}$ & $\{1,4,5\}$ & $\{1,2,4,6\}$ & $\{2,3,5,6\}$\\
 $15$ & $15/4$& $\{1,2,3\}$ & $\{1,3,4,5\}$ & $\{1,2,4,6\}$ & $\{2,3,5,6\}$\\
 $16$ &$4$& $\{1,2,3,4\}$ & $\{1,3,4,5\}$ & $\{1,2,4,6\}$ & $\{2,3,5,6\}$\\
 $17$ & $17/4$& $\{1,2,3,4\}$ & $\{1,3,4,5\}$ & $\{1,2,4,5,6\}$ & $\{2,3,5,6\}$\\
  \hline
\end{tabular}
\end{center}

With the above data placement, we plot the uplink communication load in Fig. $\text{(2a)}$ and downlink communication load in Fig. $\text{(2b)}$.
\begin{figure}[H]
	\centering
	\begin{minipage}[t]{0.5\textwidth}
		\centering
		\includegraphics[width=0.7\textwidth]{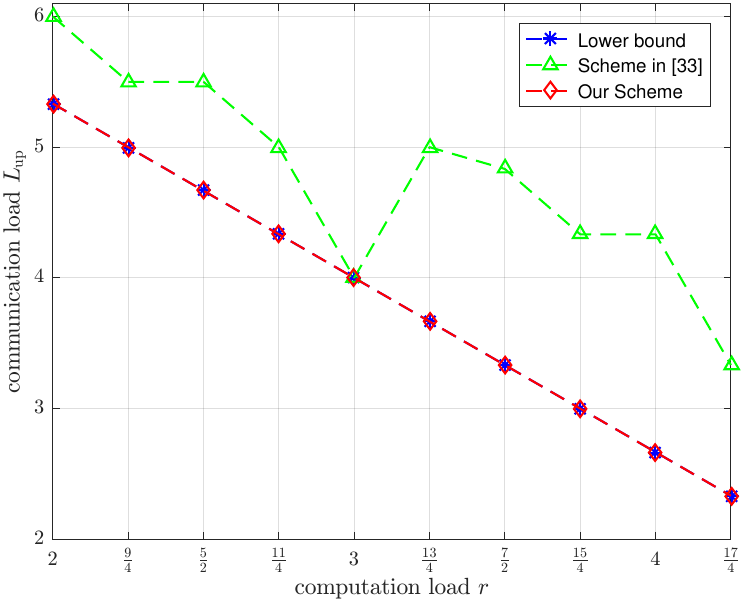} \\
		\text{(2a)} $L_{\text{up}}$ for $K=4, N=6$
		\label{fig.2a}
	\end{minipage}%
	\begin{minipage}[t]{0.5\textwidth}
		\centering
		\includegraphics[width=0.7\textwidth]{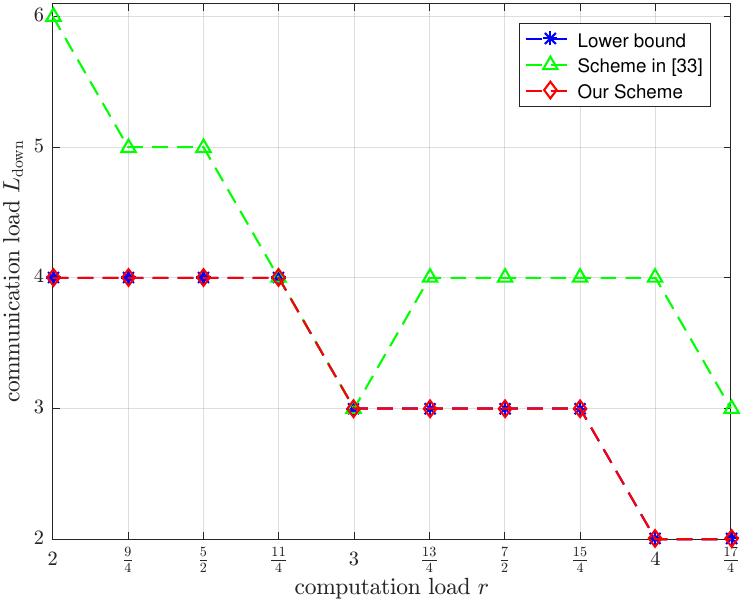} \\
		\text{(2b)}  $L_{\text{down}}$ for $K=4, N=6$
		\label{fig.2b}
	\end{minipage}
	\caption{ The $(K=4, N=6, \{Z_k\ | \ k\in[4]\})$ DMTL scheme with data placement in Table \textrm{I}}
	\label{fig.2}
\end{figure}

As illustrated in Fig. \ref{fig.2}, when $|\mathcal{Z}_k|=r=3$ with the symmetric data placement, both our scheme and the scheme in \cite{HLCMSW} achieve the optimal uplink and downlink communication loads.
For other cases, the scheme in \cite{HLCMSW} exhibits a significant gap in uplink communication load compared with our proposed scheme, even though its downlink load achieves the theoretical optimum at $r = \frac{11}{4}$.
However, our scheme always achieves the optimum for both uplink and downlink communication loads under homogeneous and heterogeneous data placement conditions.

\subsection{An Example of The Proposed Scheme in Theorem \ref{th-1}}
Let us use an example to show our scheme. Consider a $(K=4,N=5;\{\mathcal{Z}_k|k\in[4]\})$ DMTL system with the following data placement:
\begin{align}
	\label{eq-exam-caching}
	\mathcal{Z}_1=\{1,3,4\},\ \ \  \mathcal{Z}_2=\{1,2,5\},\ \ \  \mathcal{Z}_3=\{2,4\},\ \ \  \mathcal{Z}_4=\{3,5\}.
\end{align}
We have $|\mathcal{Z}_1|=|\mathcal{Z}_2|=r_1=r_2=3$ and $|\mathcal{Z}_3|=|\mathcal{Z}_4|=r_3=r_4=2$. From \eqref{eq-key-parameters} we can compute
\begin{align*}
	d_1&=d_2=5+(4-1)\times 3-(3+3+2+2)=14-10=4>0, \\
	d_3&=d_4=5+(4-1)\times 2-(3+3+2+2)=11-10=1>0,
\end{align*}and $S=d_1+d_2+d_3+d_4=8+2=10$. From \eqref{eq-combinatoric} we have
\begin{align*}
	\mathcal{P}_1=\{2,5,7,10,12,15\},\ \mathcal{P}_2=\{3,4,8,9,13,14\},\
	\mathcal{P}_3=\{1,3,5,6,8,10,11,13,15\},\ \mathcal{P}_4=\{1,2,4,6,7,9,11,12,14\}.
\end{align*}It is easy to check that {\bf Condition 1} always holds. Then we can obtain the following DMTL scheme.
\begin{itemize}
\item \textbf{Uplink Process}: Workers can locally compute their local update as follows,
	\begin{align*}
		\mathcal{V}_1=\{\bf v_1,\bf v_3,\bf v_4\},\ \mathcal{V}_2=\{\bf v_1,\bf v_2,\bf v_5\}, \
		\mathcal{V}_3=\{\bf v_2,\bf v_4\},\  \mathcal{V}_4=\{\bf v_3,\bf v_5\}.
	\end{align*}We divide each local update value into $K-1=3$ packets to obtain $N(K-1)=5\times 3=15$ packets as follows.
	\begin{align*}
		\mathbf{v}_1=\left(\begin{matrix}
			\mathbf{v}_{1,1}\\
			\mathbf{v}_{1,2}\\
			\mathbf{v}_{1,3}	
		\end{matrix}\right),
		\mathbf{v}_2=
		\left(\begin{matrix}
			\mathbf{v}_{2,1}\\
			\mathbf{v}_{2,2}\\
			\mathbf{v}_{2,3}	
		\end{matrix}\right),
		\mathbf{v}_3=\left(
		\begin{matrix}
			\mathbf{v}_{3,1}\\
			\mathbf{v}_{3,2}\\
			\mathbf{v}_{3,3}	
		\end{matrix}
		\right),
		\mathbf{v}_4=\left(
		\begin{matrix}
			\mathbf{v}_{4,1}\\
			\mathbf{v}_{4,2}\\
			\mathbf{v}_{4,3}	
		\end{matrix}
		\right),
		\mathbf{v}_5=\left(
		\begin{matrix}
			\mathbf{v}_{5,1}\\
			\mathbf{v}_{5,2}\\
			\mathbf{v}_{5,3}	
		\end{matrix}
		\right).
	\end{align*} For each integer $i\in[3]$, let
	\begin{align*}
		\widehat{\mathbf{v}}_i=\left(\begin{array}{c}
			\mathbf{v}_{1,i}\\
			\mathbf{v}_{2,i}\\
			\mathbf{v}_{3,i}\\
			\mathbf{v}_{4,i}\\
			\mathbf{v}_{5,i}
		\end{array}\right)  \ \ \text{and}\ \
		\widehat{\mathbf{v}}=
		\left(\begin{array}{c}
			\widehat{\mathbf{v}}_1\\
			\widehat{\mathbf{v}}_2\\
			\widehat{\mathbf{v}}_3
		\end{array}\right).
	\end{align*}
By coding theory, when $q>\max\{S,(K-1)N\}=15$, there always exists a $S\times (K-1)N=10\times 15$ matrix such that any sub-square is invertible. This property is called the  MDS property in coding theory. For instance, we can construct the following $S\times N(K-1)=10\times 15$ Vandermonde matrix
\begin{align*}
	\mathbf{B}=\left(
	\begin{array}{c}
		\mathbf{B}' \\\hline
		\mathbf{B}''
	\end{array}
	\right)=
	\left(
	\begin{array}{cccccccccccccccccccc}
		1&1&1&1&1&1&1&1&1&1&1&1&1&1&1\\
		w^{1}&w^{2}&w^{3}&w^{4}&w^{5}&w^{6}&w^{7}&w^{8}&w^{9}&w^{10}&w^{11}&w^{12}&w^{13}&w^{14}&w^{15}\\
		w^{2}&w^{4}&w^{6}&w^{8}&w^{10}&w^{12}&w^{14}&w^{16}&w^{18}&w^{20}&w^{22}&w^{24}&w^{26}&w^{28}&w^{30}\\
		w^{3}&w^{6}&w^{9}&w^{12}&w^{15}&w^{18}&w^{21}&w^{24}&w^{27}&w^{30}&w^{33}&w^{36}&w^{39}&w^{42}&w^{45}\\
		w^{4}&w^{8}&w^{12}&w^{16}&w^{20}&w^{24}&w^{28}&w^{32}&w^{36}&w^{40}&w^{44}&w^{48}&w^{52}&w^{56}&w^{60}\\
		w^{5}&w^{10}&w^{15}&w^{20}&w^{25}&w^{30}&w^{35}&w^{40}&w^{45}&w^{50}&w^{55}&w^{60}&w^{65}&w^{70}&w^{75}\\
		w^{6}&w^{12}&w^{18}&w^{24}&w^{30}&w^{36}&w^{42}&w^{48}&w^{54}&w^{60}&w^{66}&w^{72}&w^{78}&w^{84}&w^{90}\\
		w^{7}&w^{14}&w^{21}&w^{28}&w^{35}&w^{42}&w^{49}&w^{56}&w^{63}&w^{70}&w^{77}&w^{84}&w^{91}&w^{98}&w^{105}\\
		w^{8}&w^{16}&w^{24}&w^{32}&w^{40}&w^{48}&w^{56}&w^{64}&w^{72}&w^{80}&w^{88}&w^{96}&w^{104}&w^{112}&w^{120}\\\hline
		w^{9}&w^{18}&w^{27}&w^{36}&w^{45}&w^{54}&w^{63}&w^{72}&w^{81}&w^{90}&w^{99}&w^{108}&w^{117}&w^{126}&w^{135}
	\end{array}\right).
\end{align*}We can check that any sub-square of $\mathbf{B}$ is invertible.
The key requirement is that $\mathbf{B}$ satisfies the MDS property, enabling us to unify the uplink and downlink processes using a matrix equality in the following.

For each $\mathcal{P}_k$ where $k\in[4]$, let
\begin{align*}
	\mathbf{B}_1=\mathbf{B}([10],\mathcal{P}_1),\ \mathbf{B}_2=\mathbf{B}([10],\mathcal{P}_2),\ \mathbf{B}_3=\mathbf{B}([10],\mathcal{P}_3),\  \mathbf{B}_4=\mathbf{B}([10],\mathcal{P}_4).
\end{align*}By solving the following equations
\begin{align}
\label{eq-solution-example}
\begin{cases}
\mathbf{A}_1 \cdot \mathbf{B}_1
=\mathbf{0}_{4\times 6}\\
\mathbf{A}_2 \cdot \mathbf{B}_2
=\mathbf{0}_{4\times 6}\\
\mathbf{A}_3\cdot \mathbf{B}_3
=\mathbf{0}_{1\times 9}\\
\mathbf{A}_4\cdot \mathbf{B}_4
=\mathbf{0}_{1\times 9},
\end{cases}
\end{align}
the following solutions can be obtained.
\begin{align*}
	\mathbf{A}_1 =&\left(
	\begin{matrix}
		w^{20}&w^{25}&w^{28}&w^{17}&w^{11}&w^{22}&1&0&0&0\\
		w^{11}&w^{26}&w^{15}&w^{16}&w^{26}&w^{16}&0&1&0&0\\
		w^5&w^{28}&w^{27}&w^{16}&w^4&w^{18}&0&0&1&0\\
		w^7&w^{27}&w^{29}&w^{16}&w^{30}&w^6&0&0&0&1
	\end{matrix}\right),\\
	\mathbf{A}_2 =&\left(
	\begin{matrix}
		w^{20}&w^{15}&w^{7}&w^{6}&w^{21}&w^{12}&1&0&0&0\\
		w&w^{11}&w^{25}&w^{26}&w^{12}&w^{19}&0&1&0&0\\
		w^{8}&w^{6}&w^{4}&0&w^{8}&w^{23}&0&0&1&0\\
		w^{12}&w^{25}&w^{21}&w^{25}&w^{13}&w^{14}&0&0&0&1
	\end{matrix}\right),\\
	\mathbf{A}_3 =&\left(
	\begin{matrix}	
w^{10}~~~w^{8}~~~w^{15}~~~w^{7}~~~w^{9}~~~w~~~w^{14}~~~w^{6}~~w^{14}~~1	
	\end{matrix}
	\right),\\
	~\mathbf{A}_4 =&\left(
	\begin{matrix}
	w^{4}~~w^{20}~~~w^{5}~~~w^{23}~~~w^{4}~~~w^{26}~~w^{25}~~w^{19}~~w^{19}~~1	
	\end{matrix}\right).
\end{align*}
Then we have
\begin{align*}
	\mathbf{A}=\left(\begin{array}{c}
		\mathbf{A}_1\\
		\mathbf{A}_2\\
		\mathbf{A}_3\\
		\mathbf{A}_4
	\end{array}\right)=\left(
	\begin{array}{cccccccccc}
		w^{20}&w^{25}&w^{28}&w^{17}&w^{11}&w^{22}&1&0&0&0\\
		w^{11}&w^{26}&w^{15}&w^{16}&w^{26}&w^{16}&0&1&0&0\\
		w^{5}&w^{28}&w^{27}&w^{16}&w^{4}&w^{18}&0&0&1&0\\
		w^{7}&w^{27}&w^{29}&w^{16}&w^{30}&w^6&0&0&0&1\\\hline
		w^{20}&w^{15}&w^{7}&w^{6}&w^{21}&w^{12}&1&0&0&0\\
		w&w^{11}&w^{25}&w^{26}&w^{12}&w^{19}&0&1&0&0\\
		w^{8}&w^{6}&w^{4}&0&w^{8}&w^{23}&0&0&1&0\\
		w^{12}&w^{25}&w^{21}&w^{25}&w^{13}&w^{14}&0&0&0&1\\\hline
        w^{10}&w^{8}&w^{15}&w^{7}&w^{9}&w&w^{14}&w^{6}&w^{14}&1\\\hline
		w^{4}&w^{20}&w^{5}&w^{23}&w^{4}&w^{26}&w^{25}&w^{19}&w^{19}&1
	\end{array}\right).
\end{align*}We can check that $\mathbf{A}$ is invertible. As pointed out in Remark \ref{remark-key-idea}, %Note that
for the achievability of the general scheme, it would be challenging to prove that  $\mathbf{A}$ is always invertible given {\bf Condition 1} due to the inherent randomness in data placement.

Let $\mathbf{P}=\mathbf{A}\mathbf{B}$. We have
\begin{align*}
\mathbf{P}=&\left(\begin{array}{c}
	\mathbf{P}_1\\
	\mathbf{P}_2\\
	\mathbf{P}_3\\
	\mathbf{P}_4\\
\end{array}\right)=\left(
\begin{array}{cccc}
\mathbf{P}_{1,1} &\mathbf{P}_{1,2} &\mathbf{P}_{1,3} \\
\mathbf{P}_{2,1} &\mathbf{P}_{2,2} &\mathbf{P}_{2,3} \\
\mathbf{P}_{3,1} &\mathbf{P}_{3,2} &\mathbf{P}_{3,3}\\
\mathbf{P}_{4,1} &\mathbf{P}_{4,2} &\mathbf{P}_{4,3}\\
\end{array}
\right)\\
=&\left(
\begin{array}{ccccc|ccccc|ccccc}
	w^{16}&0&w^{18}&w^{16}&0&w^6&0&w^{25}&w^{28}&0&w^{21}&0&w^{12}&w^{20}&0\\
	w^{11}&0&w&w^{21}&0&w^{21}&0&w^{15}&w^{20}&0&w^{20}&0&w^{10}&w^{23}&0\\
	w^{20}&0&w^{21}&w^{16}&0&w^{24}&0&w^{24}&w^{22}&0&w^{27}&0&w^{25}&w^{23}&0\\
	w^4&0&w^{16}&w^2&0&w^{20}&0&w^{20}&w^{24}&0&w^{25}&0&w^{19}&w^{24}&0\\ \hline
	w^{26}&w^2&0&0&w&w^{17}&w^{22}&0&0&1&w&w^{22}&0&0&w^{10}\\
	w^{15}&w^8&0&0&w^{28}&w^{19}&1&0&0&w^{15}&w^{30}&0&0&0&w^{20}\\
	w^{19}&w^{29}&0&0&w^{3}&w^{13}&w^5&0&0&w^{15}&w^{14}&w^{10}&0&0&w^{25}\\
	w^{23}&w^{21}&0&0&w^{17}&w^{13}&w^{17}&0&0&w^{28}&w^{11}&w^3&0&0&w^{24}\\ \hline
    0&w^2&0&w^{10}&0&0&w^{28}&0&w^6&0&0&w^{15}&0&w^{25}&0\\ \hline
    0&0&1&0&w^6&0&0&w&0&w^{26}&0&0&w^6&0&w^{15}
\end{array}\right).
\end{align*}Clearly for each $y\in[3]$,
\begin{align*}
\mathbf{P}_{1,y}&=\left(
\begin{array}{ccccc}
	\ast & 0& \ast &\ast &0 \\
	\ast & 0& \ast &\ast &0 \\
	\ast & 0& \ast &\ast &0 \\
	\ast & 0& \ast &\ast &0 \\
\end{array}
\right),\ \ \ \ \ \
\mathbf{P}_{2,y}=\left(
\begin{array}{ccccc}
	\ast & \ast &0 & 0 & \ast  \\
	\ast & \ast &0 & 0 & \ast  \\
	\ast & \ast &0 & 0 & \ast  \\
	\ast & \ast &0 & 0 & \ast  \\
\end{array}
\right),\\
\mathbf{P}_{3,y}&=~\left(
\begin{array}{ccccc}
	0  &\ast & 0 &\ast &0
\end{array}
\right),\ \ \ \ \ \ \
\mathbf{P}_{4,y}=~\left(
\begin{array}{ccccc}
	0&  0& \ast &  0 & \ast
\end{array}
\right),
\end{align*} where the symbol $*$ represents an element of the finite field $\mathbb{F}_{2^5}$. This implies that every entry in column $n$ of $\mathbf{P}_{k,y}$, where $k \in [4]$ and $y \in [3]$, is $0$ if $n \in [N] \setminus \mathcal{Z}_k$. So, in the uplink process all the coded packets transmitted by the workers can be represented as follows.
\begin{align*}
\mathbf{P}\widehat{\mathbf{v}}=\left(\begin{array}{c}
	\mathbf{P}_1\\
	\mathbf{P}_2\\
	\mathbf{P}_3\\
	\mathbf{P}_4
\end{array}\right)\left(\begin{array}{c}
	\widehat{\mathbf{v}}_1\\
	\widehat{\mathbf{v}}_2\\
	\widehat{\mathbf{v}}_3
\end{array}\right)
=\left(
\begin{array}{cccc}
	\mathbf{P}_{1,1} &\mathbf{P}_{1,2} &\mathbf{P}_{1,3} \\
	\mathbf{P}_{2,1} &\mathbf{P}_{2,2} &\mathbf{P}_{2,3} \\
	\mathbf{P}_{3,1} &\mathbf{P}_{3,2} &\mathbf{P}_{3,3}\\
	\mathbf{P}_{4,1} &\mathbf{P}_{4,2} &\mathbf{P}_{4,3}
\end{array}
\right)\left(\begin{array}{c}
	\widehat{\mathbf{v}}_1\\
	\widehat{\mathbf{v}}_2\\
	\widehat{\mathbf{v}}_3
\end{array}\right)=\left(\begin{array}{c}
	\mathbf{X}_1\\
	\mathbf{X}_2\\
	\mathbf{X}_3\\
	\mathbf{X}_4
\end{array}\right).
\end{align*}Specifically,  worker $k$ transmits the coded message $\mathbf{X}_k$ to the server. From \eqref{eq-load-def} the uplink communication load is
\begin{align*}
L_{\text{up}}\triangleq\frac{\sum_{k\in[K]}T_k}{E}=\frac{4+4+1+1}{3}=\frac{10}{3}=\frac{20}{3}-\frac{10}{3}=\frac{4\times 5}{4-1}-\sum_{z\in[4]}\frac{z|\mathcal{I}_z|}{4-1}=L_{\text{up}}^\ast.
\end{align*}Thus, our scheme achieves the minimum uplink communication load.

\item \textbf{Downlink Process}:
After receiving the $4$ sent messages $\mathbf{X}_1$,  $\mathbf{X}_2$, $\mathbf{X}_3$, $\mathbf{X}_4$, i.e., $\mathbf{P}\widehat{\mathbf{v}}$, the server can obtain the coded packets
$\mathbf{A}^{-1}\mathbf{P}\widehat{\mathbf{v}}=\mathbf{B}\widehat{\mathbf{v}}$. Then the server only needs to send $\mathbf{B}'\widehat{\mathbf{v}}$ to the workers. By the MDS property each worker can decode  its required at most $(K-1)\times (N-\min_{k\in [K]}r_k)=(4-1)(5-\min_{k\in [4]}r_k)=9$ packets with help its computed local update. From \eqref{eq-load-def} the downlink communication load is
\begin{align*}
	L_{\text{down}}\triangleq \frac{T}{E}=\frac{9}{3}=3=5-2=5-\min_{k\in [4]}|\mathcal{Z}_k|=L_{\text{down}}^\ast.
\end{align*} So our downlink communication load is also minimal.
\end{itemize}
\section{Proof of Theorem \ref{th-1}}
\label{Scheme}
In this section, we employ a novel matrix decomposition to analyze the DMTL system. Specifically, our new viewpoint consists of three parts, i.e., the uplink matrix $\mathbf{P}$ characterization, the downlink matrix $\mathbf{B}$ characterization, and the existence of an invertible matrix $\mathbf{A}$ satisfying  $\mathbf{P}=\mathbf{A}\mathbf{B}$.
\subsection{Uplink Matrix $\mathbf{P}$}
\label{subsect-uplink}
Given a $(K,N;\{\mathcal{Z}_k| k\in[K]\})$ DMTL problem, in the uplink process each worker can compute its local update. Here, we regard each local update value $\mathbf{v}_n$ where $n\in[N]$ as a column vector and divide it into $K-1$ equal-length column vectors (called packets), i.e.,
	\begin{align}
		\label{eq-IV-n}
		\mathbf{v}_n=\left(\begin{array}{c}
			\mathbf{v}_{n,1}\\
			\mathbf{v}_{n,2}\\
			\vdots\\
			\mathbf{v}_{n,K-1}
		\end{array}\right).
	\end{align}
	For each $i\in[K-1]$, let $\widehat{\mathbf{v}}_i$ be the column vector generated by the packets $\mathbf{v}_{1,i}$, $\mathbf{v}_{2,i}$, $\ldots$, $\mathbf{v}_{N,i}$, i.e.,
	\begin{align}
		\label{eq-IV-in}
		\widehat{\mathbf{v}}_i=\left(\begin{array}{c}
			\mathbf{v}_{1,i}\\
			\mathbf{v}_{2,i}\\
			\vdots\\
			\mathbf{v}_{N,i}
		\end{array}\right) \ \ \text{and}\ \
		\widehat{\mathbf{v}}=
		\left(\begin{array}{c}
			\widehat{\mathbf{v}}_1\\
			\widehat{\mathbf{v}}_2\\
			\vdots\\
			\widehat{\mathbf{v}}_{K-1}
		\end{array}\right).
	\end{align}
%Let us consider the uplink communication from the viewpoint of a matrix.
Assume that the worker $k\in[K]$ sends $d_k$ coded packets of its computed local update. Then the messages can be written as
	\begin{align}
		\label{eq-uplink-commun}
		\left(\mathbf{P}_{k,1}, \mathbf{P}_{k,2},\ldots, ~\mathbf{P}_{k,K-1}\right)\left(\begin{array}{c}
			\widehat{\mathbf{v}}_1\\
			\widehat{\mathbf{v}}_2\\
			\vdots\\
			\widehat{\mathbf{v}}_{K-1}
		\end{array}\right)=\mathbf{P}_k\widehat{\mathbf{v}}
	\end{align}where $\mathbf{P}_{k,j}$ is a matrix with $d_k>0$ rows and $N$ columns over $\mathbb{F}_{q}$ for some prime power $q$ and satisfies that for any integers $l\in[d_k]$ and $n\in[N]$. The entry $\mathbf{P}_{k,j}(l,n)$ can get any element of $\mathbb{F}_{q}$ if $n\in\mathcal{Z}_k$, otherwise  $\mathbf{P}_{k,j}(l,n)=0$.
	This setting is natural since when $n\notin \mathcal{Z}_k$, the worker $k$ can not get any packet of local update value $\mathbf{v}_n$. Then all the messages sent in the uplink process can be represented as
	\begin{align}
		\label{eq-uplink-commun-detail}\left(
		\begin{array}{cccc}
			\mathbf{P}_{1,1}&\mathbf{P}_{1,2}&\cdots&\mathbf{P}_{1,K-1} \\
			\mathbf{P}_{2,1}&\mathbf{P}_{2,2}&\cdots&\mathbf{P}_{2,K-1} \\
			& & \ddots & \\
			\mathbf{P}_{K,1}&\mathbf{P}_{K,2}&\cdots&\mathbf{P}_{K,K-1}
		\end{array}
		\right)\widehat{\mathbf{v}}=\left(\begin{array}{c}
			\mathbf{P}_1\\
			\mathbf{P}_2\\
			\vdots\\
			\mathbf{P}_K
		\end{array}\right)\widehat{\mathbf{v}}=
		\mathbf{P}\widehat{\mathbf{v}}.
	\end{align} Let $S=\sum_{k\in[K]}d_k$. Then $\mathbf{P}$ has $S$ rows and $N(K-1)$ columns. In this paper, such matrix $\mathbf{P}$ is called an uplink matrix.

In the following, let us consider the condition of minimum value of $S$. Recall that each local update value is divided into $K-1$ packets. From \eqref{eq-HWSLJZ-converse} we have
\begin{align*}
S\geq (K-1)L_{\text{up}}^\ast \geq KN-\sum_{z\in[K]}z|\mathcal{I}_z|=KN-\sum_{k\in[K]}r_k.
\end{align*}If each worker $k$ sends $d_k=N+(K-1)r_k-\sum_{k'\in[K]}r_{k'}$ coded packets, the server receives exactly
\begin{align*}
S=\sum_{k\in[K]}\left(N+(K-1)r_k-\sum_{k'\in[K]}r_{k'}\right)=KN-\sum_{k\in[K]}r_k =KN-\sum_{z\in[K]}z|\mathcal{I}_z|\leq (K-1)L_{\text{up}}^\ast
\end{align*} coded packets.
Thus, we achieve the minimum uplink load $L_{\text{up}}^\ast=S/(K-1)$, which is formalized as the following result.
\begin{proposition}\rm
	\label{pro-1}
	A $(K,N;\{\mathcal{Z}_k\ | \ k\in[K]\})$ DMTL scheme has the minimum uplink communication load if each worker $k\in[K]$ sends $d_k=N+(K-1)r_k-\sum_{k'\in[K]}r_{k'}$ coded packets.
	\hfill $\square$
\end{proposition}
When {\bf Condition 1} holds, we can obtain the uplink matrix $\mathbf{P}$ with the minimum number $S$ by constructing the following matrix $\mathbf{B}$.
\subsection{Downlink Matrix $\mathbf{B}$}
\label{subsect-down}
Recall that the lower bound $L_{\text{down}}^\ast$ in  \eqref{eq-HWSLJZ-converse} is
\begin{align*}
	L_{\text{down}}^\ast  \geq N-\min_{k\in [K]}|\mathcal{Z}_k|=
	N-\min_{k\in [K]}r_k,
\end{align*}
and our uplink transmission divides each $\mathbf{v}_n$  into $K-1$ packets. We now construct the downlink matrix $\mathbf{B}$ that sends $\lambda\triangleq(K-1)(N-\min_{k\in [K]}r_k)$ coded packets, i.e., achieving the optimal $L_{\text{down}}^*= \frac{\lambda}{(K-1)}= N-\min_{k\in [K]}r_k$.   Here,  $\lambda$ is the minimum number of packets required by each worker, i.e., each worker demands at most $\lambda$ packets. %which  is exactly $\lambda\triangleq(K-1)(N-\min_{k\in [K]}r_k)$}.

It is well known that the messages sent by the server can be represented as $\mathbf{B}'\widehat{\mathbf{v}}$ where $\mathbf{B}'\in \mathbb{F}_{q}^{\lambda\times N(K-1)}$. Then each worker can always decode its demanded local update values based on its local update if $\mathbf{B}'$ has the MDS property, i.e., any sub-square of  $\mathbf{B}'$ is invertible. As pointed out in Remark \ref{remark-key-idea}, the MDS property is important in our scheme. Naturally, if we can find an appropriate matrix $\mathbf{A}\in \mathbb{F}_{q}^{S\times  S}$ and $\mathbf{B}''\in \mathbb{F}_{q}^{(S-\lambda)\times N(K-1)}$ such that
\begin{align}
	\label{eq-main-idea}
	\mathbf{P}\widehat{\mathbf{v}}=\mathbf{A}\left(
	\begin{array}{c}
		\mathbf{B}' \\
		\mathbf{B}''
	\end{array}
	\right)\widehat{\mathbf{v}}=\mathbf{A}\cdot\mathbf{B}\widehat{\mathbf{v}}
\end{align} holds,  then \emph{\eqref{eq-main-idea} can uniformly represent the uplink transmission strategy for the workers (i.e., $\mathbf{P}\widehat{\mathbf{v}}$) and the downlink transmission strategy for the server (i.e., $\mathbf{B}'\widehat{\mathbf{v}}$ of $\mathbf{A}^{-1}\mathbf{P}\widehat{\mathbf{v}}$)}. Assume that $\mathbf{B}$ has $S$ rows and satisfies the MDS property. By coding theory, the submatrix $\mathbf{B}'$ also has the MDS property. There are many studies on the MDS codes in coding theory. For instance, when $q>\max\{S, N(K-1)\}$, there always exists a matrix having the MDS property \cite{Coding}. In this paper, the matrix $\mathbf{B}$ satisfying the MDS property is called downlink matrix. There are many classic constructions of these matrices in coding theory, such as Vandermonde matrix \cite{1977The} and Cauchy matrix \cite{1989On}, etc.

\subsection{Invertible Matrix $\mathbf{A}$}
\label{subsection-invertible-A}
Given a downlink matrix $\mathbf{B}$, we only need to consider whether there exists an invertible matrix $\mathbf{A}$ such that \eqref{eq-main-idea} always holds. For each integer $k\in [K]$, recall that $\mathcal{Z}_k$ represents the index set of the data batches stored by worker $k$. This implies that the local update value can be locally computed by worker $k$ if and only if its index belongs to $\mathcal{Z}_k$. In addition, each local update value is divided into $K-1$ packets and the index of a packet uniquely corresponds to the label of a column of $\mathbf{P}$. So a packet can be locally computed by worker $k$ if and only if its index corresponds to the zero column of $\mathbf{P}$. Let $\mathcal{P}_k$ represent the set of column indices of the zero columns in $\mathbf{P}_k$, i.e., $\mathcal{P}_k = \bigcup_{i \in [0 : K-2]} \left\{ ([N] \setminus \mathcal{Z}_k) + iN \right\}$ in \eqref{eq-combinatoric}. Clearly, a packet can be locally computed by worker $k$ if and only if its index belongs to the set  $[N(K-1)]\setminus\mathcal{P}_k$.

In the following, we will show that our desired invertible matrix $\mathbf{A}$ always exists if $\mathcal{P}_k$ satisfies some condition for each integer $k\in[K]$. That is the following result.
\begin{lemma}\rm
\label{pro-2}
Our desired invertible matrix $\mathbf{A}$ always exists if {\bf Condition 1}, i.e.,  $| \bigcap_{i\in[k]}\mathcal{P}_i |\leq S-\sum_{i\in[k]}d_i$ holds for any positive integer $k\in[K]$, where $d_k$, $S$ and $\mathcal{P}_k$ are defined in \eqref{eq-key-parameters} and \eqref{eq-combinatoric} respectively.
\hfill $\square$
\end{lemma}
In order to prove Lemma \ref{pro-2}, we first use the generalized Hall's Marriage Theorem in \cite{P.C} to partition the $\mathcal{P}_k$ into $k$ disjoint subsets, denoted by $\mathcal{P}_{k,1}$, $\mathcal{P}_{k,2}$, $\ldots$, $\mathcal{P}_{k,k-1}$ and $\mathcal{P}_{k,k}$, such that $|\mathcal{P}_{k,i}|=d_i$ for each integer $i\in[k-1]$ and $|\mathcal{P}_{k,k}|=S-\sum_{i\in[k]}d_{i}$. Then, using this key property and the MDS property of $\mathbf{B}$, we can show the existence of $\mathbf{A}$. Since the proof is complex and technical, the detailed proof is included in Appendix \ref{appdix-lemma4}.

Based on the above introduction, our optimal DMTL scheme can be represented by  Algorithm 1.
\begin{algorithm}[htb]\label{algorithm}
	\caption{$(K,N=K;\{\mathcal{Z}_k|k\in[K]\})$ optimal DMTL scheme}\label{alg0}
	\begin{algorithmic}[1]
		\Procedure {Uplink}{$\mathcal{D}_1,\mathcal{D}_2,\ldots,\mathcal{D}_N$}
		\For{$k\in[K]$}
		\State $r_k\leftarrow|\mathcal{Z}_k|$
		\State Compute local model set $\mathcal{V}_k$ and split each local model into $\{\mathbf{v}_{n,i}|i\in[K-1]\}$ of equal packets
		\EndFor
		\State Obtain column vector $\widehat{\mathbf{v}}$ in \eqref{eq-IV-in}
		\For{$k\in[K]$}
		\State $d_k \gets N + (K-1)r_k - \sum_{k'\in [K]} r_{k'}$
		\State  $\mathcal{P}_k \gets \bigcup_{i\in [0 : K-2]} \left\{ ([N] \setminus \mathcal{Z}_k) + iN \right\}$
		\EndFor
		\State  $S \gets \sum_{k \in [K]} d_k$
		\State Construct a $S\times (K-1)N$ MDS matrix $\mathbf{B}$ using a Vandermonde matrix over $\mathbb{F}_q$ where $q>\max\{S, N(K-1)\}$
		\For{$k\in[K]$}
		\State  $\mathbf{B}_k \gets \mathbf{B}([S], \mathcal{P}_k)$
		  \State Solve system: $\mathbf{A}_k\mathbf{B}_k = \mathbf{0}$
		\State $\mathbf{P}_k\gets \mathbf{A}_k\mathbf{B}([S], [N(K-1)])$
		\State Worker $k$ sends $\mathbf{X}_k=\mathbf{P}_k\widehat{\mathbf{v}}$
		\EndFor
		\EndProcedure
		\Procedure{Downlink}{$\{\mathbf{A}_k\}_{k\in[K]},\{\mathbf{X}_k\}_{k\in[K]}$}
		\State $\mathbf{A} \gets (\mathbf{A}_1^\top \ \mathbf{A}_2^\top \ \cdots \ \mathbf{A}_K^\top)^\top$
		\State $\mathbf{X} \gets (\mathbf{X}_1^\top \ \mathbf{X}_2^\top \ \cdots \ \mathbf{X}_K^\top)^\top=
   (\mathbf{P}_1^\top \ \mathbf{P}_2^\top \ \cdots \ \mathbf{P}_K^\top)^\top\widehat{\mathbf{v}}  =\mathbf{P}\widehat{\mathbf{v}}$
   \State  $\lambda \gets (K-1)(N-\min_{k\in [K]}r_k)= {\rm max}_{k\in[K]} \{\mathcal{P}_k\}$
		\State Server broadcasts $\mathbf{B}'\widehat{\mathbf{v}}= (\mathbf{A}^{-1}\mathbf{X})([\lambda])=(\mathbf{A}^{-1}\mathbf{P}\widehat{\mathbf{v}})([\lambda])$ to all workers
		\State Worker $k$ decodes $\widehat{\mathbf{v}}= (\mathbf{B}'(\mathcal{P}_k,\mathcal{P}_k))^{-1}\mathbf{B}'(\mathcal{P}_k,[N(K-1)]\backslash \mathcal{P}_k)$
         \State Worker $k$ performs global update $(\omega_1,\omega_2,\cdots,\omega_K)
         =\phi(\mathbf{v}_1,\mathbf{v}_2,\cdots,\mathbf{v}_N)$ in \eqref{global update}
		\EndProcedure
	\end{algorithmic}
\end{algorithm}
\section{Conclusion}\label{CONCLUSION}
This paper presented a novel coded transmission scheme to address communication bottleneck challenges in DMTL systems. We developed a unified theoretical framework based on matrix decomposition that achieves coordinated optimization of both uplink and downlink communication strategies. Our proposed scheme demonstrates superior performance by significantly reducing the uplink and downlink communication loads. Theoretical analysis reveals when a mild condition on workers' data placement is satisfied, our scheme can simultaneously attain the theoretical lower bounds for uplink and downlink communication loads. Importantly, this optimal performance holds not only in homogeneous computing environments but also extends to various heterogeneous scenarios. Notably, the proposed scheme exhibits excellent scalability, with its core methodology being directly applicable to distributed LSC and related application scenarios.

\appendices
\section{Proof of Lemma \ref{pro-2}}
\label{appdix-lemma4}
First, let us recall some assumptions and notations. There are $N$ data subsets $\{\mathcal{D}_n\}_{n\in [N]}$  assigned to $K$ workers. $\mathcal{Z}_k$ where $k\in[K]$ represents the set of data indices stored by worker $k$ and $|\mathcal{Z}_k|=r_k$. Each local update value is divided into $K-1$ packets and each worker $k$ sends $d_k=N+(K-1)r_k-\sum_{k'\in[K]}r_{k'}$ coded packets. Let $S=\sum_{k\in[K]}d_k=KN-\sum_{k\in[K]}r_k$, which is exactly the value of $(K-1)L_{\text{up}}^\ast$. $\mathbf{B}$ is the downlink matrix of size $S\times N(K-1)$ and has the MDS property.
	
Using the above notations, let us consider the uplink communication first. The uplink messages sent by workers can be represented as in \eqref{eq-uplink-commun-detail},
	where $\mathbf{P}$ is the uplink matrix,  $\mathbf{P}_{k,y}$, where $k\in[K],y\in[K-1]$, is a $d_k\times N$ matrix over $\mathbb{F}_q$ for some prime power $q$, and  $\widehat{\mathbf{v}}$ is the local update value column vector defined in \eqref{eq-uplink-commun}.
	
	Recall that for each integer $k\in[K]$, the notation $\mathcal{P}_k$ defined in \eqref{eq-combinatoric} represents the set of column indices of the zero columns in $\mathbf{P}_k$. Define
	\begin{align}
		\label{eq-B-k}
		\mathbf{B}_k&=
		\mathbf{B}([S],\mathcal{P}_k).
	\end{align}It is easy to check that $|\mathcal{P}_k|=N(K-1)-r_k(K-1)=S-d_k$ always holds since $S=\sum_{k\in[K]}d_k=KN-\sum_{k\in[K]}r_k$. Recall that $\mathbf{B}$ has the MDS property, i.e., any square submatrix of $\mathbf{B}$ is invertible. Then, for each integer $k\in [K]$, we can always obtain a full rank matrix $\mathbf{A}_k\in \mathbb{F}_{q}^{d_{k}\times S}$ satisfying $\mathbf{A}_k\mathbf{B}_k=\mathbf{0}$. Define
	\begin{align*}
		\mathbf{A}=\left(\begin{array}{c}
			\mathbf{A}_1\\
			\mathbf{A}_2\\
			\vdots\\
			\mathbf{A}_{ K}
		\end{array}\right).
	\end{align*}
	
	In the following, we will use {\bf Condition 1}, i.e., $| \bigcap_{i\in[k]}\mathcal{P}_i |\leq S-\sum_{i\in[k]}d_i$ holds for any positive integer $k\in[K]$, to show that $\mathbf{A}$ is invertible.
	
	Let $\mathcal{A}_k$ and $\mathcal{B}_k$ denote the subspace spanned by the row vectors of the matrix $\mathbf{A}_k$ and by the column vectors of the matrix $\mathbf{B}_k$, respectively.
	Clearly, $\mathcal{A}_{k}$ is the orthogonal complement of $\mathcal{B}_{k}$. It is well known that
	\begin{align}
		\text{dim}\left(\sum_{k\in[K]}\mathcal{A}_{k}\right)\nonumber =&\sum_{i\in[K]}\text{dim}(\mathcal{A}_k)-\sum_{k\in [2 : K]}\text{dim}\left(\left(\sum_{j\in [k-1]}\mathcal{A}_{j}\right)\bigcap\mathcal{A}_k\right)\nonumber\\
		=&\sum_{i\in[K]}d_k-\sum_{k\in [2 : K]}\text{dim}\left(\left(\sum_{j\in [k-1]}\mathcal{A}_{j}\right)\bigcap\mathcal{A}_k\right)
		=S-\sum_{k\in [2 : K]}\text{dim}\left(\left(\sum_{j=1}^{k-1}\mathcal{A}_{j}\right)\bigcap\mathcal{A}_k\right).\label{eq-rank-A}
	\end{align}In the following, we only need to show $\sum_{k\in [2 : K]}\text{dim}\left(\left(\sum_{j\in [k-1]}\mathcal{A}_{j}\right)\cap\mathcal{A}_k\right)=0$.
	
	Let us first consider the case $k=2$. Then we have $\text{dim}\left(\mathcal{A}_1\cap\mathcal{A}_2\right)=\text{dim}\left((\mathcal{B}_{1}+\mathcal{B}_2)^{\bot}\right)$. By {\bf Condition 1}, we have
	\begin{align*}
		&\left|\mathcal{P}_{1}\bigcup\mathcal{P}_2\right|
		=|\mathcal{P}_{1}|+|\mathcal{P}_{2}|-|\mathcal{P}_{1}\bigcap\mathcal{P}_{2} |\geq (S-d_1)+(S-d_{2})-(S-(d_1+d_2))=S.
	\end{align*}Recall that $\mathbf{B}$ has the MDS  property. We have $\text{dim}(\mathcal{B}_{1}+\mathcal{B}_2)=S$ which implies that $\text{dim}\left(\mathcal{A}_1\cap\mathcal{A}_2\right)=0$. In the following, let us consider the case $k>2$. That is,
	\begin{align}
		&\text{dim}\left(\left(\sum_{j\in[k-1]}\mathcal{A}_{j}\right)\bigcap\mathcal{A}_{k}\right)=\text{dim}\left(\left(\bigcap_{j\in[k-1]}\mathcal{B}_{j}+\mathcal{B}_{k}\right)^\bot\right).
		\label{eq-dual-A}
	\end{align}
	It is well known that
	\begin{align}
		\text{dim}\left(\bigcap_{j\in[k-1]}\mathcal{B}_{j}+\mathcal{B}_{k}\right)
		=&\text{dim}\left(\bigcap_{j\in[k-1]}\mathcal{B}_{j}\right)+\text{dim}(\mathcal{B}_{k})-\text{dim}\left(\bigcap_{j\in[k]}\mathcal{B}_{j}\right)\label{eq-inter-dim-1}\\
		=&\left(S-\sum_{j\in[k-1]}d_j\right)+(S-d_k)-\left(S-\sum_{j\in[k]}d_j\right)\label{eq-inter-dim-2}\\
		=&S.\label{eq-value-inter-dim}
	\end{align} Here \eqref{eq-inter-dim-2}, i.e.,  $\text{dim}\left(\bigcap_{j\in[k-1]}\mathcal{B}_{j}\right)=S-\sum_{j\in[k-1]}d_j$ and $\text{dim}\left(\bigcap_{j\in[k]}\mathcal{B}_{j}\right)=S-\sum_{j\in[k]}d_j$ holds by the following result whose proof is included in Appendix \ref{appendix-Lemma-inter-dim}.
	\begin{lemma}\rm
		\label{lem-inter-dim}
		For any positive integer $k\in[K]$, we have  $\text{dim}\left(\bigcap_{i\in[k]}\mathcal{B}_{i}\right)=S-\sum_{i\in[k]}d_i$ if {\bf Condition 1} holds.
		\hfill $\square$
	\end{lemma}
	
	Substituting \eqref{eq-inter-dim-1} into \eqref{eq-dual-A} we have $\text{dim}\left(\left(\sum_{j\in[k-1]}\mathcal{A}_{j}\right)\bigcap\mathcal{A}_{k}\right)=0$ for each integer $k\geq 3$. This implies that $$\sum_{k\in [2 : K]}\text{dim}\left(\left(\sum_{j\in [k-1]}\mathcal{A}_{j}\right)\cap\mathcal{A}_k\right)=0.$$ Then $\text{dim}\left(\sum_{k\in[K]}\mathcal{A}_{k}\right)=S$ always holds.
\section{The proof of Lemma \ref{lem-inter-dim}}
\label{appendix-Lemma-inter-dim}
Let us consider any nonzero vector $\mathbf{v}\in \bigcap_{i\in[k]}\mathcal{B}_{i}$. By linear algebra, we have
$\mathbf{v}=\mathbf{B}_1\mathbf{a}_1=\mathbf{B}_2\mathbf{a}_2=\cdots=\mathbf{B}_{k-1}\mathbf{a}_{k-1}=\mathbf{B}_k\mathbf{a}_k$ where $\mathbf{a}_i$ is a coefficient column vector with length $S-d_i$ for each integer $i\in[k]$. Then the following linear system of equations can be obtained,

\begin{align*}
\begin{cases}
\mathbf{B}_1 {\bm \alpha}_1 -\mathbf{B}_k {\bm \alpha}_k = \mathbf{0} \\
\mathbf{B}_2 {\bm \alpha}_2 -\mathbf{B}_k {\bm \alpha}_k = \mathbf{0} \\
~~~~~~~~~\vdots\\
\mathbf{B}_{k-1} {\bm \alpha}_{k-1} -\mathbf{B}_k {\bm \alpha}_k = \mathbf{0}
\end{cases}
\end{align*}which can be written as
\begin{equation}\label{matrix equation}
\left(
\begin{array}{cccccc}
\mathbf{B}_1&\mathbf{0}&\cdots&\mathbf{0} &-\mathbf{B}_k \\
\mathbf{0}&\mathbf{B}_2&\cdots&\mathbf{0} &-\mathbf{B}_k \\
\vdots &\vdots   & \ddots   &\vdots & \vdots\\
\mathbf{0}&\mathbf{0}&\cdots&\mathbf{B}_{k-1} &-\mathbf{B}_k \\
\end{array}
\right)
\left(
\begin{array}{c}
{\bm \alpha}_1\\
{\bm \alpha}_2\\
\vdots\\
{\bm \alpha}_k
\end{array}
\right)=\mathbf{D}\left(
\begin{array}{c}
{\bm \alpha}_1\\
{\bm \alpha}_2\\
\vdots\\
{\bm \alpha}_k
\end{array}
\right)=\mathbf{0}.
\end{equation}

Our proof consists of two main steps: First, we will show that the rank of $\mathbf{D}$ is $(k-1)S$. This implies that the dimension of the solution space in \eqref{matrix equation} must be $(kS-\sum_{i\in[k]}d_i)-(k-1)S=S-\sum_{i\in[k]}d_i$. Secondly, we will show that the dimension of the intersection of the space $\mathcal{B}_i$ where $i\in[k]$ equals the dimension of this solution space. That is, dim$(\bigcap_{i\in[k]}\mathcal{B}_{i})=S-\sum_{i\in[k]}d_i$.
\subsection{Full Row Rank of $\mathbf{D}$}
We will use the following useful result to show our claim. The proof is included in Appendix \ref{sect-lemma-mathching}
\begin{lemma}\rm\label{lem-matching}
For each integer $k\in[k:K]$, there exist $k-1$ disjoint subsets $\mathcal{P}_{k,i}$ where $i\in[k-1]$, of $\mathcal{P}_k$ satisfying that
\begin{align*}
|\mathcal{P}_{k,i}|=d_i,\ \ \mathcal{P}_{k,i}\bigcap \mathcal{P}_i=\emptyset
\end{align*}if {\bf Condition 1} holds.
\hfill $\square$
\end{lemma}

Let $\mathcal{P}_{k,k}=\mathcal{P}\setminus(\cup_{i\in[k-1]}\mathcal{P}_{k,i})$. For each integer $i\in[k]$, define
\begin{align}
\label{eq-B-k-i}
\mathbf{B}_{k,i}=\mathbf{B}([S],\mathcal{P}_{k,i}), \ \ \ \forall i\in [k].
\end{align}From \eqref{eq-B-k} and \eqref{eq-B-k-i}, the columns of $\mathbf{B}_{k,i}$ are different from those of $\mathbf{B}_{i}$ since  $\mathcal{P}_{k,i}\cap \mathcal{P}_i=\emptyset$ where $i\in[k-1]$. In addition, it is well known that permuting the columns of a matrix does not affect its rank. So we have
\begin{align}
\label{matrix1}
\text{rank}(\mathbf{D})=&\text{rank}\left(\left(
\begin{array}{ccccccccc}
\mathbf{B}_1 & \mathbf{0} & \cdots & \mathbf{0} & -\mathbf{B}_{k,1} & -\mathbf{B}_{k,2} & \cdots & -\mathbf{B}_{k,k-1} &-\mathbf{B}_{k,k}  \\
\mathbf{0} &\mathbf{B}_2 & \cdots &\mathbf{0} & -\mathbf{B}_{k,1} & -\mathbf{B}_{k,2} & \cdots & -\mathbf{B}_{k,k-1} &-\mathbf{B}_{k,k} \\
\vdots &   \vdots   & \ddots   &\vdots & \vdots & \vdots  & \ddots  & \vdots & \vdots  \\
\mathbf{0} & \mathbf{0} &\cdots &\mathbf{B}_{k-1} &-\mathbf{B}_{k,1} & -\mathbf{B}_{k,2} & \cdots & -\mathbf{B}_{k,k-1} &-\mathbf{B}_{k,k}
\end{array}\right)\right)\nonumber\\
=&\text{rank}\left(
\left(
\begin{array}{ccccccccc}
\mathbf{B}_1  & -\mathbf{B}_{k,1} & \mathbf{0}  & -\mathbf{B}_{k,2} & \cdots & \mathbf{0}   & -\mathbf{B}_{k,k-1} &-\mathbf{B}_{k,k}  \\
\mathbf{0} & -\mathbf{B}_{k,1} &\mathbf{B}_2 & -\mathbf{B}_{k,2}&  \cdots &\mathbf{0}    & -\mathbf{B}_{k,k-1} &-\mathbf{B}_{k,k} \\
\vdots &   \vdots  &\vdots  & \vdots & \ddots     & \vdots   & \vdots & \vdots  \\
\mathbf{0} &-\mathbf{B}_{k,1}  & \mathbf{0} & -\mathbf{B}_{k,2}&\cdots  &\mathbf{B}_{k-1}  & -\mathbf{B}_{k,k-1} &-\mathbf{B}_{k,k}
\end{array}\right)
\right).
\end{align}Recall that $\mathbf{B}$ has the MDS property. We have rank$(\mathbf{B}_i,\mathbf{B}_{k,i})=S$ for each $i\in[k-1]$. This implies that
\begin{align*}
\text{rank}\left(
\left(
\begin{array}{cc}
\mathbf{B}_1  & -\mathbf{B}_{k,1}\\
\mathbf{0} & -\mathbf{B}_{k,1}\\
\vdots &   \vdots \\
\mathbf{0} &-\mathbf{B}_{k,1}
\end{array}\right)
\right)=\text{rank}\left(
\left(
\begin{array}{cc}
\mathbf{0}  & -\mathbf{B}_{k,2} \\
\mathbf{B}_2 & -\mathbf{B}_{k,2}\\
\vdots  & \vdots \\
\mathbf{0} & -\mathbf{B}_{k,2}
\end{array}\right)
\right)=\cdots=\text{rank}\left(
\left(
\begin{array}{cc}
\mathbf{0}   & -\mathbf{B}_{k,k-1}\\
\mathbf{0}    & -\mathbf{B}_{k,k-1}\\
\vdots   & \vdots\\
\mathbf{B}_{k-1}  & -\mathbf{B}_{k,k-1}
\end{array}\right)
\right)=S.
\end{align*} It is easy to check that the following submatrix
\begin{align*}
\mathbf{D}'=\left(
\begin{array}{ccccccc}
\mathbf{B}_1  & \mathbf{B}_{k,1} & \mathbf{0}  & \mathbf{B}_{k,2} & \cdots & \mathbf{0}   & \mathbf{B}_{k,k-1}\\
\mathbf{0} & \mathbf{B}_{k,1} &\mathbf{B}_2 & \mathbf{B}_{k,2}&  \cdots &\mathbf{0}    & \mathbf{B}_{k,k-1} \\
\vdots &   \vdots  &\vdots  & \vdots & \ddots     & \vdots   & \vdots  \\
\mathbf{0} &\mathbf{B}_{k,1}  & \mathbf{0} & \mathbf{B}_{k,2}&\cdots  &\mathbf{B}_{k-1}  & \mathbf{B}_{k,k-1}
\end{array}\right)
\end{align*}is a square matrix of order $(k-1)S$.

Now we aim to  show that rank$(\mathbf{D}')=(k-1)S$, which implies that rank$(\mathbf{D})=(k-1)S$. In fact, it is not easy to directly prove  rank$(\mathbf{D}')=(k-1)S$. It is well known that by finite field theory, if the rank of $\mathbf{D}'$ is exactly $(k-1)S$ if and only if the space spanned by the row vectors of the matrix $\mathbf{D}'$ has exactly $q^{(k-1)S}$ different $(k-1)S$-row vectors. So in the following, we will consider the space spanned by the row vectors of the matrix $\mathbf{D}'$.

Let $\mathbf{C}$ be a matrix whose rows are all vectors of length $(k-1)S$ over $\mathbb{F}_{q}$. That is, $\mathbf{C}$ has exactly $\mathbb{F}_{q}^{(k-1)S}$ different row vectors. Divide $\mathbf{C}$ into $k-1$ submatrices as $\mathbf{C} = \left( \mathbf{C}_1 \ \mathbf{C}_2  \ \cdots \ \mathbf{C}_{k-1} \right)$ where for each $i\in[k-1]$ $\mathbf{C}_i$ has exactly $S$ columns. Then the matrix generated by all the vectors in the space spanned by the row vectors of the matrix $\mathbf{D}'$ can be written as follows.
	\begin{align}
	&\left( \mathbf{C}_1 \ \mathbf{ C}_2 \   \cdots \  \mathbf{ C}_{k-1}\right)\left(
	\begin{array}{ccccccc}
		\mathbf{B}_1  & \mathbf{B}_{k,1} & \mathbf{0}  & \mathbf{B}_{k,2} & \cdots & \mathbf{0}   & \mathbf{B}_{k,k-1}\\
		\mathbf{0} & \mathbf{B}_{k,1} &\mathbf{B}_2 & \mathbf{B}_{k,2}&  \cdots &\mathbf{0}    & \mathbf{B}_{k,k-1} \\
		\vdots &   \vdots  &\vdots  & \vdots & \ddots     & \vdots   & \vdots  \\
		\mathbf{0} &\mathbf{B}_{k,1}  & \mathbf{0} & \mathbf{B}_{k,2}&\cdots  &\mathbf{B}_{k-1}  & \mathbf{B}_{k,k-1}
	\end{array}\right) \label{matrix-whole}\\
	=&\left( \mathbf{C}_1 \ \mathbf{ C}_2 \   \cdots \  \mathbf{ C}_{k-1}\right)\left(
	\begin{array}{ccccccc}
		\mathbf{B}_1  & \mathbf{B}_{k,1} & \mathbf{0}  & \mathbf{0} & \cdots & \mathbf{0}   & \mathbf{0}\\
		\mathbf{0} & \mathbf{B}_{k,1} &\mathbf{B}_2 & \mathbf{B}_{k,2}&  \cdots &\mathbf{0}    & \mathbf{0} \\
		\vdots &   \vdots  &\vdots  & \vdots & \ddots     & \vdots   & \vdots  \\
		\mathbf{0} &\mathbf{B}_{k,1}  & \mathbf{0} & \mathbf{0}&\cdots  &\mathbf{B}_{k-1}  & \mathbf{B}_{k,k-1}
	\end{array}\right)\label{matrix-1}\\
	+&\left( \mathbf{C}_1 \ \mathbf{ C}_2 \   \cdots \  \mathbf{ C}_{k-1}\right)\left(
	\begin{array}{ccccccc}
		\mathbf{0}  & \mathbf{0} & \mathbf{0}  & \mathbf{B}_{k,2} & \cdots & \mathbf{0}   & \mathbf{0}\\
		\mathbf{0} & \mathbf{0} &\mathbf{0} & \mathbf{0}&  \cdots &\mathbf{0}    & \mathbf{0} \\
		\vdots &   \vdots  &\vdots  & \vdots & \ddots     & \vdots   & \vdots  \\
		\mathbf{0} &\mathbf{0} & \mathbf{0} & \mathbf{0}&\cdots  &\mathbf{0}  & \mathbf{0}
	\end{array}\right)\label{matrix-k}\\
	+&\cdots\nonumber \\
	+&\left( \mathbf{C}_1 \ \mathbf{ C}_2 \   \cdots \  \mathbf{ C}_{k-1}\right)\left(
	\begin{array}{ccccccc}
		\mathbf{0}  & \mathbf{0} & \mathbf{0}  & \mathbf{0} & \cdots & \mathbf{0}   & \mathbf{0}\\
		\mathbf{0}  & \mathbf{0} & \mathbf{0}  & \mathbf{0} & \cdots & \mathbf{0}   & \mathbf{0}\\
		\mathbf{0} & \mathbf{0} &\mathbf{0} & \mathbf{0}&  \cdots &\mathbf{0}    & \mathbf{0} \\
		\vdots &   \vdots  &\vdots  & \vdots & \ddots     & \vdots   & \vdots  \\
		\mathbf{0} &\mathbf{0}  & \mathbf{0} &  \mathbf{B}_{k,2}&\cdots  &\mathbf{0}  & \mathbf{0}
	\end{array}\right)\label{matrix-2k-2}\\
	+&\cdots\nonumber \\
	+&\left( \mathbf{C}_1 \ \mathbf{ C}_2 \   \cdots \  \mathbf{ C}_{k-1}\right)\left(
	\begin{array}{ccccccc}
		\mathbf{0} & \mathbf{0} & \mathbf{0}  & \mathbf{0} & \cdots & \mathbf{0}   &  \mathbf{B}_{k,k-1}\\
		\mathbf{0} &\mathbf{0} &\mathbf{0} & \mathbf{0}&  \cdots &\mathbf{0}    & \mathbf{0} \\
		\vdots &   \vdots  &\vdots  & \vdots & \ddots     & \vdots   & \vdots  \\
		\mathbf{0} &\mathbf{0}  & \mathbf{0} &  \mathbf{0}&\cdots  &\mathbf{0} & \mathbf{0}
	\end{array}\right)\label{matrix-kk-2k}\\
	+&\cdots\nonumber \\
	+&\left( \mathbf{C}_1 \ \mathbf{ C}_2 \   \cdots \  \mathbf{ C}_{k-1}\right)\left(
	\begin{array}{ccccccc}
		\mathbf{0}  & \mathbf{0}& \mathbf{0}  & \mathbf{0} & \cdots & \mathbf{0}   &  \mathbf{0}\\
		\mathbf{0} & \mathbf{0} &\mathbf{0}& \mathbf{0}&  \cdots &\mathbf{0}    & \mathbf{0} \\
		\vdots &   \vdots  &\vdots  & \vdots & \ddots     & \vdots   & \vdots  \\
		\mathbf{0} &\mathbf{0}  & \mathbf{0} &  \mathbf{0}&\cdots  &\mathbf{0}  & \mathbf{B}_{k,k-1}\\
		\mathbf{0} &\mathbf{0}  & \mathbf{0} &  \mathbf{0}&\cdots  &\mathbf{0}  & \mathbf{0}
	\end{array}\right)\label{matrix-kk-k}
\end{align}In order to show that all the row vectors in $\mathbf{C}\mathbf{D}'$ are different, we will use the following main idea. We first show that all the row vectors in \eqref{matrix-1}, i.e.,
\begin{align}
\mathbf{D}_1=&\left( \mathbf{C}_1 \ \mathbf{C}_2  \ \cdots \ \mathbf{C}_{k-1} \right)\left(\begin{array}{ccccccc}
	\mathbf{B}_1  & \mathbf{B}_{k,1} & \mathbf{0}  & \mathbf{0} & \cdots & \mathbf{0}   & \mathbf{0}\\
	\mathbf{0} & \mathbf{B}_{k,1} &\mathbf{B}_2 & \mathbf{B}_{k,2} &  \cdots &\mathbf{0}    & \mathbf{0} \\
	\vdots &   \vdots  &\vdots  & \vdots & \ddots     & \vdots   & \vdots  \\
	\mathbf{0} &\mathbf{B}_{k,1}  & \mathbf{0} & \mathbf{0}&\cdots  &\mathbf{B}_{k-1}  & \mathbf{B}_{k,k-1}
\end{array}\right)\nonumber \\
=&\left(\mathbf{C}_1\mathbf{B}_1, \left(\sum_{i\in[k-1]}\mathbf{C}_i\right)\mathbf{B}_{k,1},
\mathbf{C}_2\mathbf{B}_2, \mathbf{C}_2\mathbf{B}_{k,2},
\cdots, \mathbf{C}_{k-1}\mathbf{B}_{k-1},  \mathbf{C}_{k-1}\mathbf{B}_{k,k-1}\right)\label{eq-2-item}
\end{align}
are different. Since all the rows of $\mathbf{D}_1$ forms a linear subspace, we only need to count the occurrence number of zero vector.
Assume that there exists a vector $\mathbf{c}=(\mathbf{c}_1,\mathbf{c}_2,\ldots,\mathbf{c}_{k-1})$, where $\mathbf{c}_i$ has length $S$ for each integer $i\in[k-1]$, satisfying that
\begin{align}\label{eq-zero}
	\left(\mathbf{c}_1\mathbf{B}_1, \left(\sum_{i\in[k-1]}\mathbf{c}_i\right)\mathbf{B}_{k,1},
	\mathbf{c}_2\mathbf{B}_2, \mathbf{c}_2\mathbf{B}_{k,2},
	\cdots, \mathbf{c}_{k-1}\mathbf{B}_{k-1},  \mathbf{c}_{k-1}\mathbf{B}_{k,k-1}\right)=(\mathbf{0},\mathbf{0},\ldots,\mathbf{0}).
\end{align}	That is, $\left(\mathbf{c}_1\mathbf{B}_1, \left(\sum_{i\in[k-1]}\mathbf{c}_i\right)\mathbf{B}_{k,1}\right)=\mathbf{0}$ and $\mathbf{c}_i(\mathbf{B}_i,\mathbf{B}_{k,i})=\mathbf{0}$ where $i\in [2,k]$. Since rank$(\mathbf{B}_i,\mathbf{B}_{k,i})=S$ where $i\in [2,k]$ i.e., $(\mathbf{B}_i,\mathbf{B}_{k,i})$ is invertible, the equation $\mathbf{c}_i(\mathbf{B}_i,\mathbf{B}_{k,i})=\mathbf{0}$ has only the zero solution.  Thus $\mathbf{c}_i=\mathbf{0}$ for $i\in [2,k]$. Then we have $\left(\mathbf{c}_1\mathbf{B}_1, \left(\sum_{i\in[k-1]}\mathbf{c}_i\right)\mathbf{B}_{k,1}\right)=\left(\mathbf{c}_1\mathbf{B}_1, \mathbf{c}_1\mathbf{B}_{k,1}\right)=\mathbf{0}$. This implies that $\mathbf{c}_1=\mathbf{0}$. Therefore, there exactly exist one solution $\mathbf{c}=(\mathbf{0},\mathbf{0},\ldots,\mathbf{0})$. Then all the row vectors in $\mathbf{C}\mathbf{D}'$ are different. Recall that $(\mathbf{B}_i,\mathbf{B}_{k,i})$ is invertible and $\mathbf{C}_i$ contains all the row vector of $\mathbb{F}_q^{S}$ for each integer $i\in[k]$. Then there exists a permutation matrix $\mathbf{Q}_1$ for $\mathbf{C}_1$ such that
\begin{align}
\label{eq-permutation-1}
\mathbf{Q}_1\mathbf{C}_1=\left(\mathbf{C}_1\mathbf{B}_1, \left(\sum_{i\in[k-1]}\mathbf{C}_i\right)\mathbf{B}_{k,1}\right).
\end{align}From \eqref{matrix-1} and \eqref{eq-permutation-1}, the matrix $\mathbf{D}_{2,1}=\mathbf{D}_1+\eqref{matrix-k}$ can be written as follows.
\begin{align}
\mathbf{D}_{2,1}=\mathbf{D}_1+\eqref{matrix-k}=\mathbf{C}\mathbf{F}_{2,1}=&\left( \mathbf{C}_1 \ \mathbf{C}_2  \ \cdots \ \mathbf{C}_{k-1} \right)\left(\begin{array}{ccccccc}
	\mathbf{B}_1  & \mathbf{B}_{k,1} & \mathbf{0}  & \mathbf{B}_{k,2} & \cdots & \mathbf{0}   & \mathbf{0}\\
	\mathbf{0} & \mathbf{B}_{k,1} &\mathbf{B}_2 & \mathbf{B}_{k,2} &  \cdots &\mathbf{0}    & \mathbf{0} \\
	\vdots &   \vdots  &\vdots  & \vdots & \ddots     & \vdots   & \vdots  \\
	\mathbf{0} &\mathbf{B}_{k,1}  & \mathbf{0} & \mathbf{0}&\cdots  &\mathbf{B}_{k-1}  & \mathbf{B}_{k,k-1}
\end{array}\right) \nonumber\\
=&\left(\mathbf{Q}_1\mathbf{C}_1,
\mathbf{C}_2\mathbf{B}_2, \mathbf{C}_2\mathbf{B}_{k,2},
\ldots, \mathbf{C}_{k-1}\mathbf{B}_{k-1}, \mathbf{C}_{k-1}\mathbf{B}_{k,k-1}\right)+\left(\mathbf{0},\mathbf{0},\mathbf{C}_1\mathbf{B}_{k,2}, \cdots \mathbf{0},\mathbf{0}\right).
\end{align}

In the following, we will show that all the row vectors in $\mathbf{D}_{2,1}$ are also distinct. Similarly, all the row vectors of $\mathbf{D}_{2,1}$ also form a linear subspace. So we only need to count the number of occurrences of the zero row vector in $\mathbf{D}_{2,1}$ too.

Let us consider the algebraic structure of $\mathbf{D}_1$ under row permutations,
\begin{align}\label{space matrix1}
	\mathbf{D}'_{1}=\begin{array}{c@{\hspace{-5pt}}l}
		\begin{array}{ccc}
			~~~~~~\overbrace{\rule{20mm}{0mm}}^{S-d_2}
			~~\overbrace{\rule{20mm}{0mm}}^{d_2}
			~~\overbrace{\rule{14mm}{0mm}}^{(k-3)S}
		\end{array} \\
		\left(\begin{array}{ccccc|c|cccccccc}
			\mathbf{c}_{1} &	0 & 0 & \cdots &0 & 0\ \ \ \ \ \cdots\ \ \ \ 0  &  \mathbf{0}  & \cdots & \mathbf{0} \\
			\mathbf{c}_{2} &	0 & 0 & \cdots &0 &0\ \ \ \ \ \cdots\ \ \ \ 0&  \mathbf{0}  & \cdots &\mathbf{0}  \\
			\vdots & \vdots     &\vdots  & \ddots &\vdots  &\vdots \ \ \ \ \  \ddots \ \ \ \ \    \vdots  & \vdots  & \ddots & \vdots\\
			\mathbf{c}_{q^S} &	0 & 0 & \cdots &0 &0\ \ \ \ \ \cdots\ \ \ \ 0&  \mathbf{0}  & \cdots & \mathbf{0}  \\\hline
			\mathbf{c}_{1} &	0 & 0 & \cdots &0 &0\ \ \ \ \ \cdots\ \ \ \ 1&  \mathbf{0}  & \cdots & \mathbf{0} \\
			\mathbf{c}_{2} &	0 & 0 & \cdots &0 &0\ \ \ \ \ \cdots\ \ \ \ 1&  \mathbf{0}  & \cdots & \mathbf{0}  \\
			\vdots & \vdots     &\vdots  & \ddots &\vdots  &\vdots \ \ \ \ \  \ddots \ \ \ \ \    \vdots  & \vdots  & \ddots & \vdots\\
			\mathbf{c}_{q^S} & 	0 & 0 & \cdots &0 &0\ \ \ \ \ \cdots\ \ \ \ 1& \mathbf{0} & \cdots & \mathbf{0}  \\\hline
			\vdots & \vdots     &\vdots  & \ddots &\vdots  &\vdots \ \ \ \ \  \ddots \ \ \ \ \    \vdots  & \vdots  & \ddots & \vdots\\
			\mathbf{c}_{1} & 0 & 0 & \cdots &0 & q-1  \cdots q-1   & \mathbf{0}  & \cdots & \mathbf{0}  \\
			\mathbf{c}_{2} &	0 & 0 & \cdots &0 & q-1	\cdots q-1   &  \mathbf{0}  & \cdots & \mathbf{0}  \\
			\vdots & \vdots     &\vdots  & \ddots&\vdots &\vdots \ \ \ \ \  \ddots \ \ \ \ \    \vdots  & \vdots  & \ddots & \vdots\\
			\mathbf{c}_{q^S} & 	0 & 0 & \cdots &0 & q-1	\cdots q-1   &\mathbf{0}  & \cdots &\mathbf{0}  \\
		\end{array}\right)
	\end{array}
\end{align}where $\mathbf{c}_{1}$, $\mathbf{c}_{2}$, $\ldots$, $\mathbf{c}_{q^S}$ are the $q^{S}$ different vectors in $\mathbf{C}_1$. Clearly $\mathbf{D}'_{1}$ has exactly $q^{S+d_2}$ different rows. For each vector $\mathbf{d}\in\mathbb{F}_q^{d_2}$, denote the solution set of $\mathbf{C}_1\mathbf{B}_{k,2}=\mathbf{d}$ by $\mathcal{G}_{\mathbf{d}}$. It is easy to check that $|\mathcal{G}_{\mathbf{d}}|=q^{S-d_2}$ and the subsets $\mathcal{G}_{\mathbf{d}}$ for all  $\mathbf{d}\in\mathbb{F}_q^{d_2}$ are the partition of $\mathbb{F}_q^{S}$ since $\mathbf{B}_{k,2}$ is a full column rank matrix.

Denote the permutation $\mathbf{Q}_1$ by the mapping $\sigma$: $\mathbb{F}_q^{S}\rightarrow \mathbb{F}_q^{S}$. Clearly, the subsets $ \sigma(\mathcal{G}_{\mathbf{d}})$ for all $\mathbf{d}\in\mathbb{F}_q^{d_2}$ are the partition of $\mathbb{F}_q^{S}$, i.e., the set $\{\mathbf{c}_{1}, \mathbf{c}_{2}, \ldots, \mathbf{c}_{q^S}\}$. Without loss of generality, we can assume that
\begin{align*}
	 \sigma(\mathcal{G}_{(0,0,\ldots,0)})&=\{\mathbf{c}_1,\mathbf{c}_2, \ldots,\mathbf{c}_{q^{S-d_2}}\},\ \
	 \sigma(\mathcal{G}_{(1,0,\ldots,0)})=\{\mathbf{c}_{q^{S-d_2}+1},\mathbf{c}_{q^{S-d_2}+2}, \ldots,\mathbf{c}_{2q^{S-d_2}}\},\\
	\cdots&\\
	 \sigma(\mathcal{G}_{(q-1,q-1,\ldots,q-1)})&=\{\mathbf{c}_{(q^{d_2}-1)q^{S-d_2}+1},\mathbf{c}_{(q^{d_2}-1)q^{S-d_2}+2}, \ldots,\mathbf{c}_{q^{S}}\}
\end{align*} Then in the matrix in \eqref{matrix-k}, we also take the same row labels of $\mathbf{D}_1$ and order them in the same way as $\mathbf{D}'_{1}$ to obtain a new matrix $\mathbf{F}$. It is sufficient to count the occurrence number of the zero vector in $\mathbf{D}'_1+\mathbf{F}$. We can obtain the following submatrix where all the entries in the column labels from $[S+1:(k-1)S]$ are $0$.
\begin{align}\label{space matrix1}
	\mathbf{D}'_{2,1}=\begin{array}{c@{\hspace{-5pt}}l}
		\begin{array}{ccc}
			~~~~~~~~~~~~~~~~~~~~~\overbrace{\rule{19mm}{0mm}}^{S-d_2}
			~~\overbrace{\rule{14mm}{0mm}}^{d_2}
			~~\overbrace{\rule{14mm}{0mm}}^{(k-3)S}
		\end{array} \\
		\left(\begin{array}{ccccc|ccc|ccccc}
			\mathbf{c}_{1} &0 & 0 & \cdots & 0 & 0 &\cdots & 0   &  \mathbf{0}  & \cdots & \mathbf{0} \\
			\mathbf{c}_{2} &0 & 0 & \cdots & 0 &0 &\cdots & 0 &\mathbf{0}  & \cdots &\mathbf{0}  \\
			\vdots & \vdots     &\vdots  & \ddots &\vdots  &\vdots   &\ddots    & \vdots  & \vdots  & \ddots & \vdots\\
			\mathbf{c}_{q^{S-d_2}} &0 & 0 & \cdots& 0 &0 &\cdots &0 &\mathbf{0}  & \cdots & \mathbf{0}  \\\hline
			\mathbf{c}_{q^{S-d_2}+1} &0 & 0 & \cdots& 0 &0 &\cdots & 0&  \mathbf{0}  & \cdots & \mathbf{0} \\
			\mathbf{c}_{q^{S-d_2}+2} &0 & 0 & \cdots& 0 &0 &\cdots &0&  \mathbf{0}  & \cdots & \mathbf{0}  \\
			\vdots & \vdots     &\vdots  & \ddots &\vdots  &\vdots  &\ddots  &\vdots  & \vdots  & \ddots & \vdots\\
			\mathbf{c}_{2q^{S-d_2}} & 0 & 0 & \cdots& 0 &0 &\cdots & 0& \mathbf{0} & \cdots & \mathbf{0}  \\\hline
			\vdots & \vdots     &\vdots  & \ddots &\vdots  &\vdots  &\ddots   &  \vdots  & \vdots  & \ddots & \vdots\\\hline
			\mathbf{c}_{(q^{d_2}-1)q^{S-d_2}+1} & 0 & 0 & \cdots& 0 & 0 &\cdots& 0& \mathbf{0}  & \cdots & \mathbf{0}  \\
			\mathbf{c}_{(q^{d_2}-1)q^{S-d_2}+2} &0 & 0 & \cdots& 0 &0 &\cdots & 0&  \mathbf{0}  & \cdots & \mathbf{0}  \\
			\vdots & \vdots     &\vdots  & \ddots &\vdots  &\vdots   &\ddots    &\vdots  & \vdots  & \ddots & \vdots\\
			\mathbf{c}_{q^S} & 0 & 0 & \cdots& 0 & 0\ &\cdots & 0&\mathbf{0}  & \cdots &\mathbf{0}  \\
		\end{array}\right).
	\end{array}
\end{align}Clearly, the zero vector in $\mathbb{F}_q^{(k-1)S}$ occurs in $\mathbf{D}_{2,1}$ exactly once. Therefore, all rows of $\mathbf{D}_{2,1}$ are distinct.

Similar to the discussions of $\mathbf{D}_1$ and $\mathbf{D}_{2,1}$, we can also show that all the row vectors of the following matrices
\begin{align*}
\mathbf{D}_{2,k}=\mathbf{D}_{2,1}+\eqref{matrix-2k-2}, \ \ldots,\ \mathbf{D}_{k,1}=\mathbf{D}_{k-1,k-1}+\eqref{matrix-kk-2k}, \
\ldots, \ \mathbf{D}_{k,k-2}=\mathbf{D}_{k,k-3}+\eqref{matrix-kk-k}=\mathbf{C}\mathbf{D}'	
\end{align*} are distinct. This implies that $\mathbf{D}'$ has full row rank. Then $\mathbf{D}$ has full row rank. So, the dimension of the solution space in \eqref{matrix equation} is $(kS-\sum_{i\in[k]}d_i)-(k-1)S=S-\sum_{i\in[k]}d_i$.

\subsection{Proof of dim$(\bigcap_{i\in[k]}\mathcal{B}_{i})=S-\sum_{i\in[k]}d_i$}
In the following, we will first show that the solution space of \eqref{matrix equation} is determined by all the possible vector ${\bm \alpha}_1$. Suppose there exist two solution vectors for \eqref{matrix equation}, denoted by
\begin{align*}
\begin{pmatrix} {\bm \alpha}_1 \\ {\bm \alpha}_2 \\ \vdots \\ {\bm \alpha}_k \end{pmatrix}\ \ \text{and}\ \ \
\begin{pmatrix} {\bm \alpha}_1 \\ {\bm \alpha}_2' \\ \vdots \\ {\bm \alpha}_k' \end{pmatrix},	
\end{align*} respectively that satisfy
\begin{align*}
\mathbf{B}_1{\bm \alpha}_1 = \mathbf{B}_2{\bm \alpha}_2 = \mathbf{B}_2{\bm \alpha}_2'= \cdots = \mathbf{B}_{k-1}{\bm \alpha}_{k-1} =  \mathbf{B}_{k-1}{\bm \alpha}_{k-1}' = \mathbf{B}_k{\bm \alpha}_k= \mathbf{B}_k{\bm \alpha}_k'.	
\end{align*} This implies that $\mathbf{B}_i({\bm \alpha}_i - {\bm \alpha}_i') = \mathbf{0}$ for $i \in [2 : k]$. Then ${\bm \alpha}_i = {\bm \alpha}_i'$ holds for $i \in [2 : k]$ since $\mathbf{B}_i\mathbf{x} = \mathbf{0}$ has only the zero solution by the hypothesis that $\mathbf{B}_i$ is full column rank.

It is well known that all the possible vectors ${\bm \alpha}_1$ form a linear subspace, denoted by $\mathcal{C}_1$, since the solution space for \eqref{matrix equation} is a linear subspace. So the dimension of $\mathcal{C}_1$ is the same as the dimension of solution space for \eqref{matrix equation}, i.e., dim$(\mathcal{C}_1)=S-\sum_{i\in[k]}d_i$. Since $\mathbf{B}_1$ is a full column rank matrix, $\mathbf{B}_1\mathcal{C}_1$ is also a linear subspace and has dimension $S-\sum_{i\in[k]}d_i$. In addition, we have showed dim$(\bigcap_{i\in[k]} \mathcal{B}_i)=\text{dim}(\mathcal{C}_1)$. So, ${\rm{dim}}\left(\bigcap_{i\in[k]}\mathcal{B}_{i}\right)=\dim(\mathcal{C}_1)=S-\sum_{i\in[k]}d_i$ always holds.

\section{The proof of Lemma \ref{lem-matching}}
\label{sect-lemma-mathching}
For each $i\in[k-1]$, let $\mathcal{A}_i=\mathcal{P}_k\setminus\mathcal{P}_i=\mathcal{P}_k\setminus(\mathcal{P}_k\bigcap\mathcal{P}_i)$. In addition, we have $|\mathcal{A}_i|\geq d_i$ since  $|\mathcal{P}_k\bigcap\mathcal{P}_i|\leq S-(d_k+d_i)$. For each $i\in[k-1]$, let $\mathcal{S}_i=\{x_{i,j}| \ 1\leq j\leq d_i\}$ be the set of $d_i$ different labels. For any $i\neq j\in[k-1]$, $\mathcal{S}_i\bigcap\mathcal{S}_j=\emptyset$. Let $\mathcal{X}=\bigcup_{i\in[k-1]}\mathcal{S}_i$ and  $\mathcal{Y}=\bigcup_{i\in[k-1]}\mathcal{A}_i$. Define a bipartite graph $G=(\mathcal{X},\mathcal{Y},\mathcal{E})$ where $\mathcal{E}=\{(y,x_{i,j})\ |\ y\in\mathcal{A}_i, i\in [k-1],~j\in[d_i] \}$, i.e., $y \in \mathcal{A}_i$ connects to all vertices $x_{i,j} \in \mathcal{S}_i$ where $i\in [k-1]$.
For any subset $\mathcal{X}'$ of $\mathcal{X}$, define
\begin{align*}
\Gamma(\mathcal{X}') = \{ y \in \mathcal{Y} \mid \exists \  x_{i,k} \in \mathcal{X}' \text{ such that } (y,x_{i,k}) \in \mathcal{E}, \forall~ i\in [k-1]\},
\end{align*}
i.e., $\Gamma(\mathcal{X}')$ is the neighbor set of $\mathcal{X}'$ in $\mathcal{Y}$. A matching in graph $G$ is a set of edges between left and right partitions where no two edges share a common endpoint. By the following result in \cite{P.C}, we will show that there exists a matching  that covers every vertex in $\mathcal{X}$.
\begin{lemma}[Hall’s Marriage Theorem, \cite{P.C}]\label{Hall-marriage}
Given a bipartite graph $G=(\mathcal{X},\mathcal{Y},\mathcal{E})$ with $|\mathcal{X}|\leq |\mathcal{Y}|$, if $|\Gamma(\mathcal{X}')|\geq |\mathcal{X}|$ holds for any vertex subset $\mathcal{X}'$ of $\mathcal{X}$, there exists a matching that covers every vertex in $\mathcal{X}$.
\hfill $\square$
\end{lemma}
In order to show  $|\Gamma(\mathcal{X}')|\geq |\mathcal{X}'|$, the following notation is useful. For any subset $\mathcal{X}'$ of $\mathcal{X}$, define
\begin{align*}
	\mathcal{I}=\{i\in [k-1]|~\exists~ j\in[d_i], x_{i,j}\in \mathcal{X}'\}.
\end{align*}

For any $i\in \mathcal{I}$, let $l_i=|\{j\in [d_i]|~x_{i,j}\in \mathcal{X}'\}|$. By our hypothesis, we have $|\mathcal{X}'|=\sum_{i\in \mathcal{I}}l_i\leq \sum_{i\in\mathcal{I}}d_i$. By the above notation, we have
\begin{align*}
|\Gamma(\mathcal{X}')|=&|\bigcup_{i\in \mathcal{I}}\mathcal{A}_i|=|\bigcup_{i\in \mathcal{I}}(\mathcal{P}_k\setminus\mathcal{P}_i)| \\
=&|\mathcal{P}_k\setminus\bigcap_{i\in\mathcal{I}}\mathcal{P}_i|=|\mathcal{P}_k|-\left| \left(\bigcap_{i\in\mathcal{I}}\mathcal{P}_i\right) \bigcap\mathcal{P}_k \right| \\
\geq&(S-d_k)-(S-(\sum_{i\in\mathcal{I}}d_i+d_k))\\
=&\sum_{i\in\mathcal{I}}d_i\geq \sum_{i\in \mathcal{I}}l_i=|\mathcal{X}'|.
\end{align*}
By Lemma \ref{Hall-marriage}, we only need to show that for any vertex set $\mathcal{X}'\subset\mathcal{X}$, $G$ has a matching covering all the vertices in $\mathcal{X}$.
That is, there is a matching from $\mathcal{S}_i$ to $\mathcal{P}_{k,i}$, where each $\mathcal{P}_{k,i} \subseteq \mathcal{A}_i$, $i\in[k-1]$.
So there exist $k-1$ disjoint subsets $\mathcal{P}_{k,i}$ of $\mathcal{P}_k$,  and  $|\mathcal{P}_{k,i}|=d_i$, $i\in[k-1]$. Since $\mathcal{A}_i=\mathcal{P}_k\setminus\mathcal{P}_i$, we have $\mathcal{P}_{k,i}\bigcap \mathcal{P}_i=\emptyset$   for each integer $i\in[k-1]$. Then the proof is completed.
\bibliography{references}
\end{document}